\documentclass[preprint]{aastex}
\usepackage{graphicx}
\begin{document}
\title{Infrared Spectra of the Subluminous Type Ia Supernova 1999by}

\shorttitle{Infrared Spectra of SN1999by}
\shortauthors{H\"oflich et al.}
  
\author{Peter H\"oflich}
\affil{Department of Astronomy, The University of Texas at Austin, 
TX 78712 Austin, USA\altaffilmark{1}}
\altaffiltext{1}{{\tt pah@hej1.as.utexas.edu}}
\author{Christopher L. Gerardy \& Robert A. Fesen}
\affil{6127 Wilder Laboratory, Physics \& Astronomy Department \\
       Dartmouth College, Hanover, NH 03755, USA} 
\and
\author{Shoko Sakai}
\affil{Division of Astronomy and Astrophysics, UCLA, Los Angeles, CA 90095}

\begin{abstract}

Near-infrared (NIR) spectra of the subluminous Type~Ia supernova SN~1999by are
presented which cover the time evolution from about 4 days before to 2 weeks
after maximum light. Analysis of these data was accomplished through the 
construction of an extended set of delayed detonation (DD) models covering the 
entire range of normal to subluminous SNe~Ia.  The explosion, light 
curves (LC), and the time evolution of the synthetic spectra were calculated 
self-consistently for each model with the only free parameters being the initial
structure of the white dwarf (WD) and the description of the nuclear burning 
front during the explosion. From these, one model was selected for 
SN~1999by by matching the synthetic and observed optical light 
curves, principly the rapid brightness decline.  DD models require a minimum
amount of burning during the deflagration phase which implies a lower limit for the $^{56}Ni$ mass of about $0.1 M_\odot$ and consequently a lower limit for the
SN brightness. The models which best match the optical light curve of SN~1999by 
were those with a $^{56}Ni$ production close to this theoretical minimum.  The 
data are consistent with little or no interstellar reddening ($E(B-V) \leq 
0.12^m$) and we derive a distance or $11 \pm 2.5$ Mpc for SN~1999by in 
agreement with other estimates.  

Without any modification, the synthetic spectra from this subluminous model 
match reasonably well the observed IR spectra taken on May 6, May 10, May 16 and
May 24, 1999. These dates correspond roughly to $-4$~d, 0~d, and 6~d and 14~d 
after maximum light.  Prior to maximum, the NIR spectra of SN~1999by are 
dominated by products of explosive carbon burning (O, Mg), and Si. Spectra taken
after maximum light are dominated by products of incomplete Si burning.
Unlike the behavior of normal Type~Ia SNe, lines from iron-group elements only
begin to show up in our last spectrum taken about two weeks after maximum light.
The implied distribution of elements in velocity space agrees well with the
DD model predictions for a subluminous SN~Ia.  Regardless of the explosion
model, the long duration of the phases dominated by layers of explosive carbon
and oxygen burning argues that SN~1999by was the explosion of a white dwarf
at or near the Chandrasekhar mass.

The good agreement between the observations and the models without fine-tuning
a large number of free parameters suggests that DD models are a good description
of at least subluminous Type~Ia SNe.  Pure deflagration scenarios or mergers
are unlikely and helium-triggered explosions can be ruled out.  However,
problems for DD models still remain, as the data seem to be at odds with recent
3-D models of the deflagration phase which predict significant
mixing of the inner layers of the white dwarf prior to detonation.
Possible solutions include the effects of rapid rotation on the propagation of
nuclear flames during the explosive phase of burning, or extensive burning of 
carbon just prior to the runaway.

\end{abstract}

\keywords{supernovae -  individual: SN1999by}
 
\section{Introduction}

On April 30, SN~1999by was independently discovered at about $15^m$ by 
\citet{arbor99}, and \citet{papenkova99}. Images with the Katzman Automatic 
Imaging Telescope \citep{treffers97,li99} of the same field provide an 
upper limit of $19.3^m$ on April 25.2 UT. The supernova (SN) was found in the Sb
galaxy NGC~2841 which has been the host of three previous supernovae: 
SN~1912A, SN~1957A, and SN~1972R \citep{papenkova99}. Based on optical spectra, 
SN~1999by was identified as a Type~Ia SN \citep{gerardy99}.  Shortly 
thereafter, \citet{garnavich99} reported that SN~1999by showed stronger than 
normal \ion{Si}{2} 5800~\AA\ absorption and depressed flux near 4000~\AA, 
suggesting that SN~1999by would be a significantly subluminous Type~Ia event.  

According to \citet{bonanos99}, SN~1999by reached a maximum light of 
$m_B=13.80^m \pm 0.02 ^m$ on UT
1999 May 10.5, and a maximum in the $V$ band of $m_V=13.36^m \pm 0.02 ^m$.  
The Lyon/Meudon Extragalactic Database 
(LEDA)\footnote{http://www-obs.univ-lyon1.fr/leda/home\_leda.html}, gives the
heliocentric radial velocity of NGC~2841, corrected for Virgo infall, as  
$811.545$ km~s$^{-1}$.  Using $H_0=65$ km~s$^{-1}$~Mpc$^{-1}$ puts the 
distance to NGC~2841 at 12.5 Mpc with a distance modulus of $30.5^m$. Using 
this distance modulus and the photometry of \citet{bonanos99}, the absolute peak
magnitudes of SN~1999by are $M_B=-16.68^m$ and $M_V=-17.12^m$. Thus, SN~1999by 
is underluminous by roughly 2.5 magnitudes as compared to a typical 
Type~Ia supernova (SN~Ia).  

Detailed analysis of the 
optical light curves (LCs) by \citet{toth00} confirm these basic results.  
The light curve of SN~1999by shows a very steep post-maximum brightness decline 
$M_V(\Delta M_{\Delta t=15d})$ of $1.35^m$ to $1.45^m$.  Based on detailed fits 
of the LCs, \citet{toth00} find the interstellar reddening to be
$E(B-V) \leq 0.1^m$ which is in agreement with the values for galactic reddening
given by \citet[$E(B-V)=0.015^m$]{schlegel98}, and 
\citet[$E(B-V) \approx 0^m$]{burstein82}. Recently, \citet{garnavich} provided 
detailed optical LCs and spectra and found values consistent with 
previous data.

These measurements imply that SN~1999by was as underluminous as SN~1991bg, the 
prototypical example of the subluminous Type~Ia subclass 
\citep{filippenko92,leibundgut93}. Other members include SN~1992K 
\citep{hamuy96a,hamuy96b}, SN~1997cn \citep{turatto98}, and SN~1998de 
\citep{modjaz00}.  Some defining characteristics of the subclass are
rapidly declining light curves ($M_B(\Delta M_{\Delta t=15d}) \simeq 1.9^m$), peak
magnitudes fainter than normal by 2--3 mag, and redder colors at maximum light 
($(B-V)_{\rm max}$ $\simeq$ 0.4--0.5$^m$).  One of the currently active areas in
SNe~Ia research concerns the nature of these subluminous events. 
%REVISION START
Theoretical interpretations of subluminous SN~Ia include all three
types of explosion mechanism: the centrally triggered detonation of a 
sub-Chandrasekhar mass WD, deflagration and delayed detonations of massive WD
and two merging WDs (see sect. 3 \& Woosley \& Weaver, 1994 ; H\"oflich et al.,
1995; Nugent et al., 1995, Milne et al. 1999).
The possibility has also been raised that SN1991bg-like SNe~Ia should not be 
classified with other SNe~Ia at all as they may arise from different 
progenitors.
%REVISION STOP

SN~1999by is one of the best observed SNe~Ia with data superior to that of 
SN~1991bg.  In addition to the studies of optical light curves and 
spectra mentioned above, detailed polarization spectra of SN~1999by have been 
obtained and analyzed \citep{howell01}. Whereas `normal' SNe~Ia tend to show
little or no polarization \citep[e.g.][]{wang97}, this supernova was 
significantly polarized, up to 0.7\%, indicating an overall asphericity of 
$\approx 20\%$.  This result suggests that there may be a connection between 
the observed asphericity and the subluminosity in SNe~Ia.  

In recent years it has become apparent that infrared spectroscopic observations
can be used as valuable tools to determine the chemical structure of SN~Ia 
envelopes. For instance, near-infrared spectra can be used to locate the 
boundaries between explosive carbon and oxygen burning, or between complete and 
incomplete silicon burning by measuring \ion{Mg}{2}, \ion{Ca}{2} and iron-group 
lines \citep{wheeler98}.  However, the difficulties involved in obtaining high 
quality infrared spectra of supernovae has limited IR studies to a small handful
of objects.  
The study of SN~1994D by \citet{meikle96} is the only one to include both
pre- and post-maximum spectra.  Since most of their spectra was of the 
1--1.3~\micron\ region, \citet{meikle96} concentrated on determining the origin 
of a feature at 1.05~\micron. They concluded that this feature 
might be due to either \ion{He}{1} 1.083~\micron\ or \ion{Mg}{2}~1.0926 \micron,
but found difficulties with both identifications.  Detailed modeling of 
SN~1994D and SN~1986G was able to reproduce the basic NIR spectral 
features and their evolution \citep{hoflich97,wheeler98}.  
In particular, the identification of the 1.05 \micron\ feature in SN~1994D as 
due to \ion{Mg}{2} could be established 
(recently confirmed by \citealt{lentz01}), and a broad feature, 
appearing between $\approx 1.5$ and 1.9 \micron\ was identified as a blend 
of iron-group elements.   

%REVISION START
We note that the nature of subluminous SNe~Ia is important for the use of 
SNe~Ia as 
distance indicators.  In particular, the calibration of the brightness decline 
relation depends critically on subluminous SNe~Ia because they significantly
increase the required baseline observations. From this perspective, it may turn 
out to be critical to determine whether all SNe~Ia form a homogeneous class of 
objects or not.  However, current limitations of model calculations will not 
allow us to improve the accuracy of the current estimates for the absolute 
distance to SN~1999by, and the possible implications of our results on the use 
of SNe~Ia for cosmology is not the subject of this paper.
%REVISION STOP

In this paper, we present near-infrared spectra of SN~1999by covering the 
time evolution of the supernova from about four days before to two weeks 
after maximum light.  Detailed models of the SN explosion based on a
delayed detonation (DD) scenario are used to analyze the data.
The explosion models, light curves, and synthetic spectra are calculated 
in a self-consistent manner. Given the initial structure of the progenitor 
and description of the nuclear burning front, the light curves and spectra
are calculated from the explosion model without any further tuning.
The observations and data reduction are discussed in 
\S 2.  In \S 3, general properties of the explosion models, light curves and 
spectra are discussed for normal and subluminous SNe~Ia within the DD scenario. 
In \S 4, a specific model for this SN is chosen by matching the properties of 
the predicted light curve to those observed in optical studies of SN~1999by.  
Synthetic optical and IR spectra for this model are then presented and the 
latter are compared to the observed IR spectra.  Detailed line identifications 
for the infrared features are provided.  The effects on the predicted spectra of
large scale mixing during the deflagration phase also examined.  In \S 5,
we discuss the implications for progenitors and alternative explosion 
mechanisms.  We close in \S 6 with a final discussion of the results for 
SN 1999by, and put our findings
in context with other SNe~Ia and different explosion scenarios.
 
\section {Observations}
 
Low-dispersion (R $\approx$ 700), near-infrared spectra of SN~1999by 
from 1.0--2.4~\micron\ were obtained with the 2.4m 
Hiltner Telescope at MDM Observatory during the nights of 6--10 May 1999.
The data were collected using TIFKAM (a.k.a.\ ONIS), a high-throughput
infrared imager and spectrograph with an ALLADIN 512 $\times$ 1024 InSb
detector. This instrument can be operated with standard \textit{J}, \textit{H},
and \textit{K} filters for broadband imaging, or with a variety of grisms,
blocking filters, and an east-west oriented $0\farcs 6$ slit for low
($R \approx 700$) and moderate ($R \approx 1400$) resolution spectroscopy.

SN~1999by was observed with three different spectroscopic setups, which covered 
the 0.96--1.80~\micron, 1.2--2.2~\micron, and 1.95--2.4~\micron\ wavelength 
regions.  The observations were broken into multiple 600~s exposures. Between 
each exposure, the supernova light was dithered along the slit to minimize the 
effect of detector defects and provide first-order background subtraction.  
Total exposure times vary from spectrum to spectrum and are listed in the log of
NIR observations in Table~\ref{nirlog}.

Wavelength calibration of the spectra was achieved by observing neon, argon, 
and xenon arc lamps.  The spectra were corrected for telluric absorption by
observing A~stars and early G~dwarfs at similar airmass, chosen from the Bright 
Star Catalog \citep{BSC}.  Applying the procedure described by
\citet{hanson98}, stellar features were removed from
the G~dwarf spectra by dividing by a normalized solar spectrum
\citep{solar1, solar2}\footnote{NSO/Kitt Peak FTS data used
here were produced by NSF/NOAO.}. The results were used to correct for 
telluric absorption in the A~stars.  Hydrogen features in the corrected A~star 
spectra were removed from the raw A~star spectra and the results were used to 
correct the target data for telluric absorption. [For further discussion of this
procedure see \citealt{hanson98,HCR96}, and references therein.] The 
instrumental response was calibrated by matching the continuum of the A~star 
telluric standards to the stellar atmosphere models of \citet{kurucz93}.

The first three nights (6-8 May) were photometric, and \textit{J}, 
\textit{H}, and \textit{K} broadband images of SN~1999by and \citet{persson} 
photometric standards. The resulting photometry is listed in Table~\ref{nirlog}
and was used to set the flux levels of the corresponding spectra.  A square 
bandpass was assumed and the flux levels thus 
attained are believed accurate to $\sim$ 20--30\%.  For May 9 and 10 no
flux information was available, but the observed spectral bands had large
overlaps, and the relative flux levels were set by matching the
data in the overlapping regions. 

Near-infrared spectra of a star $29\farcs 5$ north and 
$11\farcs 7$ west of SN~1999by \citep{papenkova99} were also obtained 
and reduced in the same manner.  This star was then used as a local 
relative spectrophotometric standard to reduce spectroscopic data from 
later observing runs.  Spectra of SN~1999by covering 1.0--1.8~\micron\ 
were collected on five more nights during the following two weeks
(16 May, 18 May, 20 May, 21 May and 24 May). The first three were 
obtained by P. Martini and A. Steed using TIFKAM on the 
2.4m telescope at MDM, and the last two by S. Sakai using TIFKAM on 
the 2.1m telescope at the Kitt Peak National Observatory.  

\begin{figure}
\includegraphics[width=15.2cm,angle=0]{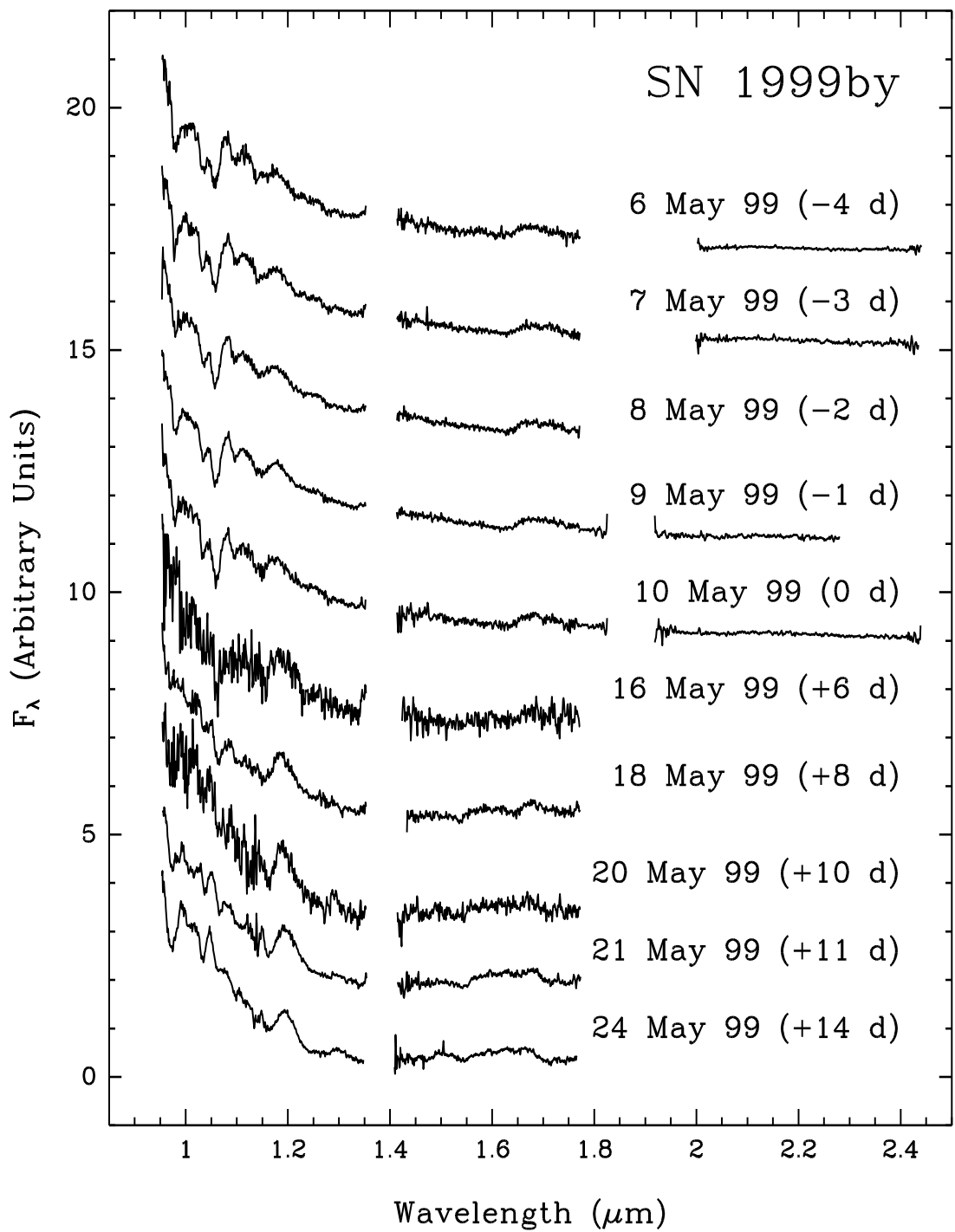}
\caption{Near-infrared spectral evolution of SN1999by.  Epochs are relative
to the date of \textit{V}$_{\rm Max}$, May 10, 1999.  For clairity, the spectra
have been shifted vertically, and regions of very low S/N due to telluric 
absorption have been omitted.}
\label{nirspec}
\end{figure}

Figure~\ref{nirspec}
shows a plot of all the NIR spectra.  Since absolute fluxes were only available
for the first three spectra, the data are presented in arbitrary flux units, and
they have been shifted vertically for clarity.  Regions of very low S/N due to
strong telluric absorption have been omitted.  The epochs listed are relative
to 10 May 1999, the date of $V_{Max}$ given by \citet{bonanos99}.  Like all the
subsequent plots, the data have been corrected to the 638 km~s$^{-1}$ redshift 
of NGC~2841 \citep{devaucouleurs91}.

\section {Models for the Explosion, Light Curves, and Spectra}

There is general agreement that SNe~Ia result from some process involving the 
combustion of a degenerate white dwarf \citep{hoyle60}. Within this general 
picture, three classes of models have been considered: (1) An explosion of a 
CO-WD, with mass close to the Chandrasekhar limit, which accretes mass through 
Roche-lobe overflow from an evolved companion star \citep{whelan73}. The 
explosion is triggered by compressional heating near the WD center.  (2) An 
explosion of a rotating configuration formed from the merging of two low-mass 
WDs, caused by the loss of angular momentum through gravitational radiation
\citep{webbink84,iben84,paczynski85}. (3) Explosion
of a low mass CO-WD triggered by the detonation of a helium layer 
\citep{nomoto80,woosley80,woosley86}. 

Only the first two models appear to be 
viable for normal Type~Ia SNe as the third, the sub-Chandrasekhar WD model, has been ruled out on the basis of predicted light curves and spectra 
\citep{hoflich96b,nugent97}.  Still, theoretical interpretations of subluminous 
SN~Ia include all three types of explosion mechanism
\citep{woosley94,hkw95,nugent95,milne99}.  The
possibility has also been raised that the subluminous SN~1991bg-like supernovae 
may arise from different progenitors and should not be classified
with other SNe~Ia at all.

Within $M_{Ch}$ models, it is believed that the burning front starts as a 
subsonic deflagration.  However, the time evolution of the burning front is 
still an open question. That is, whether the deflagration front burns 
through the entire WD \citep{nomoto84}, or alternatively, transitions into 
a supersonic detonation mode as suggested in the delayed detonation (DD) model
\citep{khokhlov91,woosley94,yamaoka92}. DD models have been found to reproduce 
the optical and infrared light curves and spectra of `typical' SNe~Ia reasonably
well (\citealt{hoflich95}, H95 hereafter; 
\citealt{hk96,hoflich96b,fisher98,nugent97,wheeler98,lentz01}).  

In addition, 
DD models provide a natural explanation for the brightness decline relation 
$M (\Delta M_{\Delta t=15d})$ observed in the light curves of SNe~Ia 
\citep{phillips87,hamuy95,hamuy96b,suntzeff99}.
This is a consequence of the temperature 
dependence of the opacity and the lack of a dependence of the explosion energy 
on the amount of $^{56}Ni$ produced 
(\citealt{hmk93,khokhlov93,hoflich96b,mazzali01}).  Thus, within
the delayed detonation scenario, both normal and 
subluminous SNe~Ia can be explained as variants of a single phenomenon 
(H95, \citealt{hkw95}, \citealt{umeda99}).  This makes the DD model an
attractive scenario and it is used as the basis for our present analysis.

In our models, the explosions LCs and spectra are all calculated
self-consistently.  Given the initial structure of
the progenitor and a description of the nuclear burning front, the light curves
and spectra are calculated directly from the explosion model without any
additional parameters. Since the predicted observables follow directly from
the supernova model, this approach provides a direct link between the
observations and the explosion physics and progenitor properties.

Finally, we note that detailed analyses of observed SN~Ia spectra and light 
curves indicate that mergers \citep{benz90} and pure deflagration SNe (such as
the W7 model) could contribute to the population of bright SNe~Ia 
(\citealt{hk96}, hereafter HK96; \citealt{hatano00}). On the other hand, it is 
not obvious that subluminous SNe~Ia and the brightness decline relation can be 
understood within pure deflagration models.  In particular, the amount of 
burning, and consequently the explosion energy, will be strongly correlated with
$^{56}Ni$ production. Since no detailed studies for deflagration models have 
been performed, these models cannot be ruled out but, as we discuss in \S 5, the
IR spectra of SN~1999by are hard to reconcile with a pure deflagration.

\subsection{Numerical Methods}

\subsubsection{The Explosion}
Despite recent progress in our understanding of nuclear burning fronts, current
3-D models are not sufficiently evolved to allow for a consistent treatment of
the burning front throughout all phases
\citep[see also \S 4.3 \& \S 5]{khokhlov01}. Thus detailed models rely on
parameterized descriptions guided by 3-D results.  Within spherical models, the
transition from deflagration to detonation can be conveniently parameterized by 
a density $\rho_{tr}$ which has been found to be the dominating factor for the 
determining chemical structure, light curves and spectra.

We have calculated explosion models using a one-dimensional radiation-hydro code
(HK96) that solves the hydrodynamical equations explicitly by the piecewise 
parabolic method \citep{collela84}.  Nuclear burning is taken into account 
using an extended network of 606 isotopes from n,p to $^{74}Kr$
\citep[and references therein]{thieleman96}. The propagation of the nuclear 
burning front is given by the velocity of sound behind the burning front in the 
case of a detonation wave, and in a parameterized form during the deflagration 
phase calibrated by detailed 3-D calculations \citep[e.g.][]{khokhlov01}.  We 
use the parameterization as described in \citet{dominguez00}.  For a 
deflagration front at distance $r_{burn}$ from the center, we assume that the 
burning velocity is given by 
$v_{\rm burn}=max(v_{t}, v_{l})$, where $v_{l}$ and $v_{t}$
are the laminar and turbulent velocities with 
$$ v_{t}= 0.2 ~\sqrt{\alpha_{T} ~ g~L_f},~
\eqno{[1]} $$
 with~
$$\alpha_T ={(\alpha-1)/( \alpha +1})$$ ~and ~
$$\alpha ={\rho^+(r_{\rm burn})/ \rho^-(r_{\rm burn})}.$$
\noindent
Here $\alpha _T$ is the Atwood number, $L_f$ is the characteristic
length scale, and $\rho^+$ and $\rho^-$ are the densities in front of
and behind the burning front, respectively. The quantities $\alpha$ and $L_f$ 
are directly taken from the hydrodynamical model at the location of the burning 
front and we take $L_f=r_{\rm burn(t)}$. The transition density is treated as a 
free parameter.

\subsubsection{The Light Curves}

 From these explosion models, the subsequent expansion, bolometric and
broad band light curves (LC) are calculated following the method described by 
\citet{hoflich98}, and references therein. The LC-code is the same used for the 
explosion except that $\gamma$-ray transport is included via a Monte Carlo 
scheme and nuclear burning is neglected. In order to allow a more consistent 
treatment of the expansion, we solve the time-dependent, frequency-averaged 
radiation moment equations. The frequency-averaged variable Eddington factors
and mean opacities are calculated by solving the frequency-dependent transport 
equations. About one thousand frequencies (in one hundred frequency groups) and
about nine hundred depth points are used.  At each time step, we use T(r) to 
determine the Eddington factors and mean opacities by solving the 
frequency-dependent radiation transport equation in a co-moving frame and 
integrate to obtain the frequency-averaged quantities.  The averaged opacities 
have been calculated assuming local thermodynamic equilibrium (LTE).  Both the 
monochromatic and mean opacities are calculated in the narrow line limit.
Scattering, photon redistribution, and thermalization terms used in the light 
curve opacity calculations are taken into account. 

In previous works (e.g. \citet{hk96}), the
photon redistribution and thermalization terms have been calibrated for a sample
of spectra using the formalism of the equivalent two level approach (H95). For 
increased consistency, we use the same equations and atomic models but solve the
rate equations simultaneously with the light curves calculation at about every 
100$^{th}$ time step, at the expense of some simplifications in the NLTE-part 
compared to H95.  For the opacities we use the narrow line limit, and for the 
radiation fields we use the solution of the monochromatic radiation transport 
using $\approx 1000$ frequency groups.

\subsubsection {Spectral Calculations}

Our non-LTE code (H95, and references therein) solves the relativistic
radiation transport equations in a co-moving frame.  The spectra are computed 
for various epochs using the chemical, density, and luminosity structure and 
$\gamma$-ray deposition resulting from the light curve coder. This provides a 
tight coupling between the explosion model and the radiative transfer.  The 
effects of instantaneous energy deposition by $\gamma$-rays, the stored energy 
(in the thermal bath and in ionization) and the energy loss due to the adiabatic
expansion are taken into account. Bound-bound, bound-free and free-free 
opacities are included in the radiation transport which has been discretized by 
about $2 \times 10^4$ frequencies and 97 radial points. 

The radiation transport 
equations are solved consistently with the statistical equations and ionization 
due to $\gamma$-radiation for the most important elements and ions. Typically, 
between 27 to 137 bound levels are taken into account.  We use \ion{C}{1} 
(27/123/242), \ion{O}{1} (43/129/431), \ion{Mg}{2} (20/60/153),
\ion{Si}{2} (35/212/506), \ion{Ca}{2} (41/195/742), \ion{Ti}{2} (62/75/592), 
\ion{Fe}{2} (137/3120/7293), \ion{Co}{2} (84/1355/5396), 
\ion{Ni}{2} (71/865/3064) where the first, second and third numbers in brackets
denote the number of levels, bound-bound transitions, and number of discrete 
lines for the radiation transport. The latter number is larger because nearby 
levels within multiplets have been merged for the rates. 

The neighboring 
ionization stages have been approximated by simplified atomic models restricted 
to a few NLTE levels + LTE levels. The energy levels and cross sections of 
bound-bound transitions are taken from \citet{kurucz93} starting at the ground 
state. The bound-free cross sections are taken from TOPBASE 0.5 as implemented 
in Stra\ss burg \citep{mendoza93}. Collisional transitions are treated
in the `classical' hydrogen-like approximation \citep{mihalas78} that relates 
the radiative to the collisional gf-values. All form factors are set to 1.

About 10$^{6}$ additional lines are included (out of a line list of 
$4 \times 10^7$) assuming LTE-level populations. The scattering, photon 
redistribution, and thermalization terms are computed with an 
equivalent-two-level formalism that is calibrated using NLTE models. 

\subsection {Modeling Results}               
 
Spherical dynamical explosions, light curves, and spectra are calculated for 
both normal bright and subluminous SNe~Ia. Due to the one-dimensional 
nature of the models, the moment of the transition to a detonation is a free 
parameter. The moment of deflagration-to-detonation transition (DDT) is 
conveniently parameterized by introducing the transition density, 
$\rho_{\rm tr}$, at which DDT happens.
 
Within the DD scenario, the free model parameters are: 1) The chemical structure
of the exploding WD, 2) Its central density $\rho_c$ at the time of the 
explosion, 3) The description of the deflagration front, and 4) The density 
$\rho_{tr}$ at which the transition from deflagration to detonation occurs. 
%REVISION START
Note that the density structure of a WD only weakly depends on the temperature
because it is highly degenerate. The central density $\rho_c$ depends mainly
on the history of accretion.
%REVISION STOP
Model parameters have been chosen for the progenitor and $\rho_c$ which allow 
us to reproduce light curves and spectra of `typical' Type Ia supernovae.

In all models, the structure of the exploding C/O white dwarf is based on a 
model star with 5 M$_{\sun}$ at the main sequence and solar metallicity. 
Through accretion, this core has been grown close to the Chandrasekhar limit 
(see Model 5p0y23z22 in \citealt{dominguez01}).  At the time of the explosion of
the WD, its central density is 2.0$\times 10^9$ g~cm$^{-3}$ and its mass is 
close to 1.37$M_\odot$. The transition density $\rho_{tr}$ has been identified 
as the main factor which determines the $^{56}Ni$ production and, thus the 
brightness of a SNe~Ia (H95; \citealt{hkw95}; \citealt{iwamoto99}). The
transition density $\rho_{tr}$ from deflagration to detonation is varied 
between 8 and 27 $\times 10^6$ g~cm$^{-3}$ to span the entire range of 
brightness in SNe~Ia. The models are identified by {\sl 5p0z22.ext} where 
{\sl ext} is the value of $\rho_{tr}$ in $10^6$ g~cm$^{-3}$. Some of the basic 
quantities are given in Table~\ref{modelprop}.

\subsubsection{Explosion Models}

Here we will restrict our discussion to the basic features of DD models 
which are relevant for the spectral analysis of SN~1999by. For a more detailed 
discussion of the hydrodynamical evolution of DD-models, see \citet{khokhlov91},
H95, and references therein. 

\begin{figure}
\includegraphics[width=12.cm,angle=360]{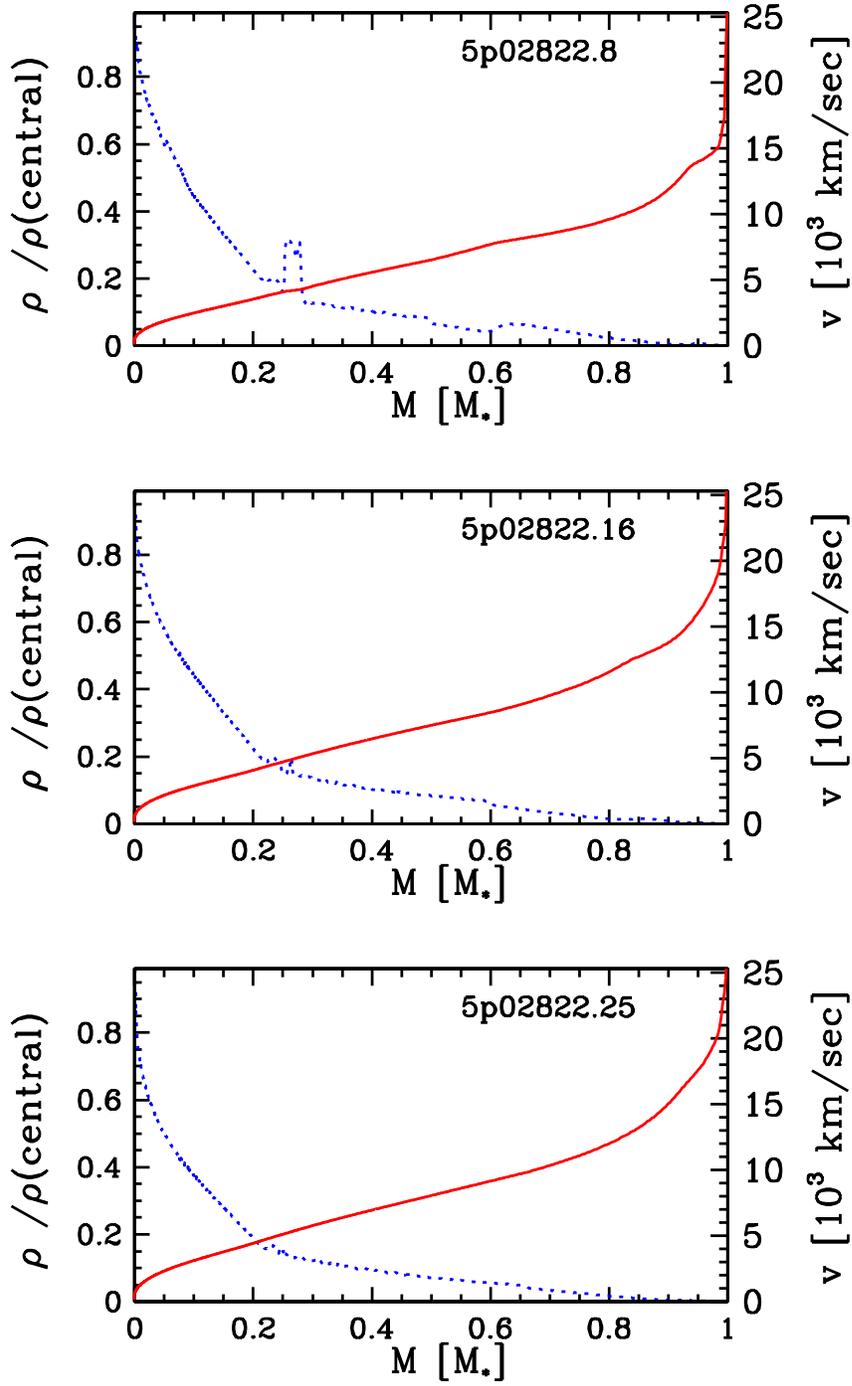}
\caption{Density (blue, dotted) and velocity (red, solid) as a function of the 
mass for models with $\rho_{tr}=$ 8, 16 and 25 $\times 10^6 g/cm^{-3}$ 
(from top to bottom).} 
\label{density}              
\end{figure}

\begin{figure}
\includegraphics[width=10.2cm,angle=180]{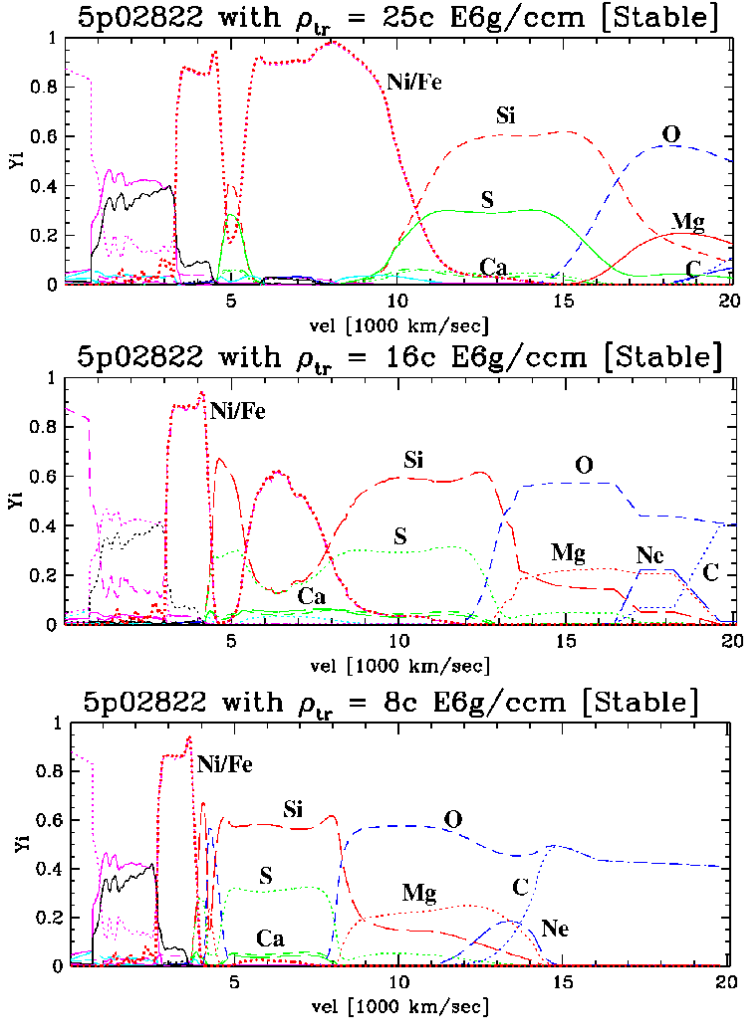}
\caption{Abundances of stable isotopes as a function of the expansion velocity 
for models with  $\rho_{tr}=$ 8, 16 and 25 $\times 10^6 g/cm^{-3}$ 
(from top to bottom). In addition, $^{56}Ni$ is given.  The curves with the 
highest abundance close to the center correspond to $^{54}Fe$, $^{58}Ni$ and 
$^{56}Fe$.} 
\label{abundance}
\end{figure}

During the deflagration phase, the WD is lifted, starts to expand, and after 
burning about 0.22 to 0.28 $M_\odot$, the transition to a detonation is 
triggered.  Figures \ref{density} and \ref{abundance} show the
density, velocity and chemical structures for representative 
models after a homologous expansion has been achieved. The final velocity and 
density structures are rather similar. The expansion velocities decrease 
slightly with $\rho_{tr}$ because a significant amount of oxygen remains 
unburned and did not contribute to the energy production. In 5p0z22.8, oxygen 
remained unburned in the outer $\approx 0.4 M_\sun $.                  
Overall, the density is decreasing with radius because burning took place 
throughout the entire WD except in the very outer layers. In contrast, pure 
deflagration model such as W7 \citep{nomoto84}, pulsating delayed detonation 
\citep{hkw95}, and merger models \citep{khokhlov93} show a significant density 
bump at the boundary between burned and unburned layers.
     
Iron group elements are produced in the inner layers where density and 
temperature stay high for a sufficient time during the explosion, whereas 
intermediate mass elements are produced in the layers above where nuclear 
statistical equilibrium (NSE) has no time to set in. In the outer zones, 
products of explosive oxygen (e.g. Si, S) and explosive carbon burning 
(O, Mg, Ne) and some Si are seen. Only a very thin layer of unburned C/O
remains.
 
For layers with final velocities $v\leq 3000$~km~s$^{-1}$, the densities in the
early stages of the explosion are sufficiently high for electron capture. This 
results in the production of $^{54,56}$Fe \& $^{58}$Ni whereas in the outer 
layers only $^{56}$Ni is produced. We employ the same description for the 
deflagration front in all models and, consequently, the very inner structures 
are almost identical in all models.  A small dip in the Ni abundance is present 
at the DDT because the time scales for burning to NSE are comparable to the 
expansion time scales at densities of $\approx 10^7$ g~cm$^{-3}$.  This feature 
can be expected to be smeared out in multi-dimensional calculations 
\citep{livne99}.

After the DDT, a detonation front develops and the burning of fuel is triggered 
by compression of unburned material.  Consequently, burning takes place under 
higher densities compared to a deflagration. In general, additional material is 
burned up to NSE and this extends the layers of $^{56}$Ni for all 
$\rho_{tr} \ge 1.2 \times 10^7$ g~cm$^{-3}$. For models with smaller 
$\rho_{tr}$, burning only proceeds up to Si despite the compression. This 
increases the production of intermediate mass elements (e.g. Si, S) at the 
expense of $^{56}$Ni. In our models, the efficiency for $^{56}$Ni production 
drops rapidly as a function of decreasing $\rho_{tr}$ 
(see Table~\ref{modelprop}).  We note that for $\rho_{tr} \leq 6 \times 10^{6}$ 
g~cm$^{-3}$, nuclear burning stops all together for $\rho < \rho_{th}$ and, 
unlike any observed SN, one would see a very slowly expanding C/O envelope 
heated by little $^{56}$Ni.
 
There exists a rather strict lower limit for the absolute 
brightness of SNe~Ia within the Chandrasekhar mass models. This can be 
understood as follows. The absolute brightness of a SNe~Ia is mainly determined 
by the $^{56}Ni$. In general, $^{56}Ni$ is produced both during the deflagration
and the detonation phase. For the most subluminous SNe~Ia however, no $^{56}Ni$ 
is produced during the detonation. Within the $M_{Ch}$ scenario and DD-models 
in particular, we need to burn a minimum of $\approx 0.2 M_\odot$ during the 
deflagration to achieve the required `lift/pre-expansion' of the WD. During the 
deflagration phase, iron group elements are produced. The fraction of 
$^{56}Ni$ depends on the amount of material which undergoes electron capture, 
i.e. on the central density and, to a smaller extent, on the description of the 
burning front (\citealt{brachwitz01};  
\citealt{dominguez01}). For a model with $\rho_c = 2 \times 10^9$ g~cm$^{-3}$, 
we find a lower limit of $M(^{56}Ni)\approx 0.09 M_\odot$.  Recently, 
\citet{brachwitz01} studied the influence of $\rho_{c}$ between 1.5 to 
$8 \times 10^9$ g~cm$^{-3}$ on the nuclear burning for similar DD-models with a 
high transition density which may describe `normal' SN~Ia. In those 
models, the $^{56}Ni$ produced during the deflagration phase varies by 
$\leq 25 \%$ which provides an estimate for the possible variation in the 
minimum $^{56}Ni $ production. We conclude that there exists a lower limit of 
$M(^{56}Ni) \approx 0.1 M_{\sun} \pm 25 \%$ within DD-models.

%REVISION START
In principle, 3-D effects may change the lower limit for the $^{56}Ni$ 
production.  However, all current 3-D calculations show that rising blobs are 
formed at high densities which undergo complete Si-burning because the burning 
time scales to Ni shorter than the hydro time scales.  In the case of a DDT or 
a transition to a phase of very fast deflagration, the amount of
burning will likely be similar to the 1-D case.  Thus, we do not 
expect a significant change in the minimum ${56}Ni$ production.
%REVISION END

\subsubsection{General properties of theoretical light curves}
 
Optical light curves and spectra of SN~Ia are not the subject of this paper but
we used their general properties for selecting a model for SN1999by.      
 
Within the DD scenario, both normal and subluminous SNe~Ia can be produced
(H95, \citealt{hkw95}, HKW95 hereafter; \citealt{umeda99}). The
amount of $^{56}$Ni ($M_{^{56}Ni}$) depends primarily on $\rho_{tr}$ (H95; 
HKW95; \citealt{umeda99}) and to a much lesser extent on the assumed value of 
the deflagration speed, initial central density of the WD, and initial chemical 
composition (ratio of carbon to oxygen). Models with smaller transition density 
give less nickel and hence both lower peak luminosity and lower temperatures 
(HKW95, \citealt{umeda99}). In DDs, almost the entire WD is burned and the 
total production of nuclear energy is almost constant. This and the dominance of
$\rho_{tr}$ for determining the $^{56}Ni$ production are the basis of why, to 
first approximation, SNe~Ia appear to be a one-parameter family. 

The observed 
$M_V(\Delta M_{\Delta t=15d})$ relation can be understood as an opacity 
effect \citep{hoflich96b}, namely, the consequence of rapidly dropping opacities
at low temperatures \citep{hmk93,khokhlov93,mazzali01}. Less Ni means lower
temperature and consequently, reduced mean opacities because the emissivity is 
shifted from the UV toward longer wavelengths with less line blocking.  A more 
rapidly decreasing photosphere causes a faster release of the stored energy and,
as a consequence, steeper declining LCs with decreasing brightness.  The DD 
models give a natural and physically well-motivated origin of the 
$M_V(\Delta M_{\Delta t=15d})$ relation of SNe~Ia 
(\citealt{hamuy95,hamuy96b,phillips87,suntzeff99}).
 
\begin{figure}
\includegraphics[width=10.2cm,angle=270]{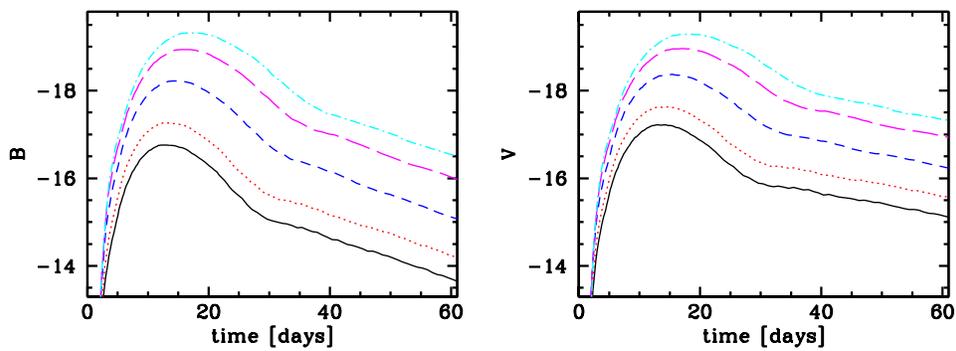}
\caption{\textit{B} (left) and \textit{V} (right) light curves for 
models with 
$ \rho_{tr} $ = 8, 12, 16, 20 and 25 $\times 10^{6}g/cm^3$ (from bottom to 
top).}
\label{lcs}
\end{figure}

\begin{figure}
\includegraphics[width=8.2cm,angle=270]{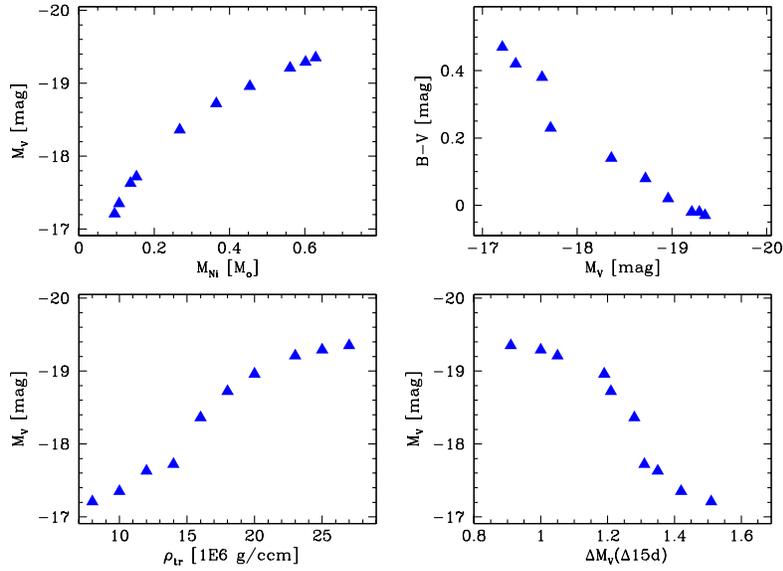}
\caption{Maximum brightness $M_V$ as a function of the $^{56}Ni$ mass (upper
left), $\rho_{tr}$ (lower left), and $M_V(\Delta M_{\Delta t=15d})$ (lower 
right), and $(B-V)$ as a function of $M_V$ (upper right).  For all panels except
the lower right, the points correspond to models with $\rho_{tr}$ of 8, 10, 12, 
14, 16, 18, 20, 23, 25 and 27~$\times 10^{6}g/cm^3$ from left to right.  In the
lower right panel, this order is reversed.}
\label{dm15}
\end{figure}

The broad band \textit{B} and \textit{V} light curves are shown in 
Figure~\ref{lcs}. Some general properties are given in Table~\ref{modelprop} and
Figure~\ref{dm15}. We note that the decline of the \textit{B} and \textit{V} LC 
during the early $^{56}Co$ tail is slightly smaller for the most 
subluminous models (e.g. 5p0z22.08) compared to the intermediate models 
(e.g. 5p0z22.16) because, at day 40, the $\gamma $-rays are almost fully trapped
in the former whereas, for the latter, the escape probability of $\gamma $-rays 
is already increasing (HKW95). With varying $\rho_{tr}$, we find rise times 
between 13.9 to 19.0 days, and the maximum brightness $M_V$ spans a range 
$-17.21^m$ and $-19.35^m$ with $B-V$ between $+0.47^m$ to $-0.03^m$.  The value
of $M_V$ is primarily determined by the transition density. Variations in the 
progenitors (main sequence mass and metallicity) or the central density at the 
point of ignition causes an spread around $M_V$ of $\approx0.2~...~0.3^m$ 
\citep{hoflich98,dominguez01}. As discussed above, the brightness decline 
relation $M_V(\Delta M_{\Delta t=15d})$ can be understood as an opacity effect 
or, more precisely, it is due to the drop of the opacity at low temperatures. 
Therefore, subluminous models are redder and the decline is steeper. In our 
series of models, $M_V(\Delta M_{\Delta t=15d})$ ranges between $1.45^m$ and
$0.91^m$. However, $M_V(\Delta M_{\Delta t=15d})$ is not a strictly linear 
relation (Fig. \ref{dm15}). For bright SNe~Ia 
($-19.35^m \leq M_V \leq -18.72^m$), called {\sl Branch-normals} (\citep{branch96,branch99}),
 the linear 
decline is relatively flat, followed by a steeper decline toward the 
subluminous models. For normal-luminosity models, the photospheric temperature 
is well above the critical value at which the opacity drops 
($\approx 7000 ~... 8000 K$; \citealt{hmk93})
whereas the opacity drops rapidly in subluminous models soon after maximum 
light.  

\section{Analysis of SN1999by} 

\subsection{Selection of the SN1999by Model} 
            
Detailed B and V light curves of SN~1999by have been collected and published by 
\citet{toth00} and optical spectra around maximum light are available from
observations at McDonald Observatory \citep{howell01}. Recently, SN~1999by 
spectra and LCs have been provided by \citet{garnavich}. Within our series of 
models, we selected the most suitable model based on the optical photometry.
Unfortunately, there is a gap in the observed data starting about one week
after maximum light, a phase which is most suitable for selecting the
appropriate models. Thus, as the main discriminator, we use the complete LCs as 
reconstructed by \citet{toth00}.

\begin{figure}
\includegraphics[width=9.2cm,angle=270]{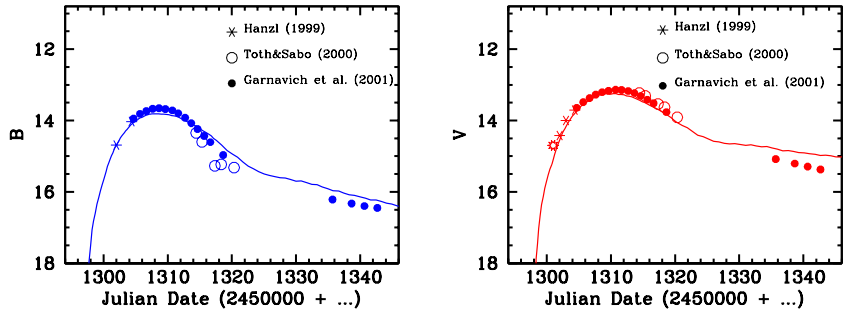}
\caption{
Comparison of the observed \textit{B} (left) and \textit{V} (right) light curves
of SN1999by with the predicted light curves of model 5p0z22.08.}
\label{lc.obs}
\end{figure}

SN~1999by showed a decline rate $M_V(\Delta M_{\Delta t=15d})$ between 
$ 1.35^m$ and $1.45^m$ (\citealt{toth00}) clearly ruling out all but the most 
subluminous models
with transition densities of $8$ and $10 \times 10^6$ g cm$^{-3}$ and maximum 
brightness of $-17.21^m$ and $-17.35^m$, respectively (Table~\ref{modelprop}).
Both models are consistent with the early light curves given in \citet{toth00} i
and \citet{garnavich} (Fig.\ref{lc.obs}).  From the models and the early LC, the
time of the explosion can be placed around April 27 $\pm$ 1~d, 1999. 
The model post-maximum LC in B is less reliable because it depends 
very sensitively on the size and time evolution of line blanketing and the 
thermalization.  $M_B(\Delta M_{\Delta t=15d})$ is $1.64^m$ and $1.73^m$ for 
5p0z22.10 and 5p0z22.8, respectively. This compares reasonably well with 
$M_B(\Delta M_{\Delta t=15d}) = 1.87^m$ observed for SN1999by \citep{howell01}.
In the LCs (Fig. \ref{lc.obs}), discrepancies of about $0.3 ^m$ show up after 
the time gap in the observations.  At these times, the envelope of a subluminous
SN~Ia becomes transparent, and strong emission features emerge (see below). 
Thus these discrepancies are likely a consequence of discretization errors due 
to the use use of only 1000 frequency for the LC calculations.
 
According to \citet{bonanos99}, $B-V = 0.44 \pm 0.04^m$ near maximum light 
on May 10, 1999. \citet{garnavich} give $B-V = 0.50 \pm 0.03^m$. We find 
$B-V = 0.47^m$ and $0.42^m$ for 5p0z22.8 and 5p0z22.10, respectively. In 
previous studies, we find intrinsic uncertainties in B-V to 
be $\approx 0.05^m$ for our models at maximum light \citep{hoflich95}. Thus, the
interstellar 
reddening can be expected to be $E_{B-V} \leq 0.11^m$ consistent with the values
for our galaxy given by \citet{burstein82} and \citet{schlegel98} and the 
estimates for the 
total extinction derived by \citet{toth00}. For our models, if we assume 
$m_V = 13.10 \pm 0.05^m$, we then derive a distance modulus of 
$M_V-m_V = 30.39^m \pm 0.12^m$ ($-0$ to 0.35$^m$). The first and second 
error-terms originate from the observational uncertainties plus the brightness 
range of the models and the reddening correction, respectively.  >From our 
models, we can derive a distance to NGC 2841 of $ 11 \pm 2.5$~Mpc and
$12 \pm 1$~Mpc if we include and neglect interstellar reddening, respectively.
Both from observations \citep{schlegel98} and our models, very small 
reddening is preferred. Following \citet{schlegel98}, we adopt $E(B-V)=0.015^m$
for all comparisons with the observations.
  
\begin{figure}
\includegraphics[width=13.2cm,angle=360]{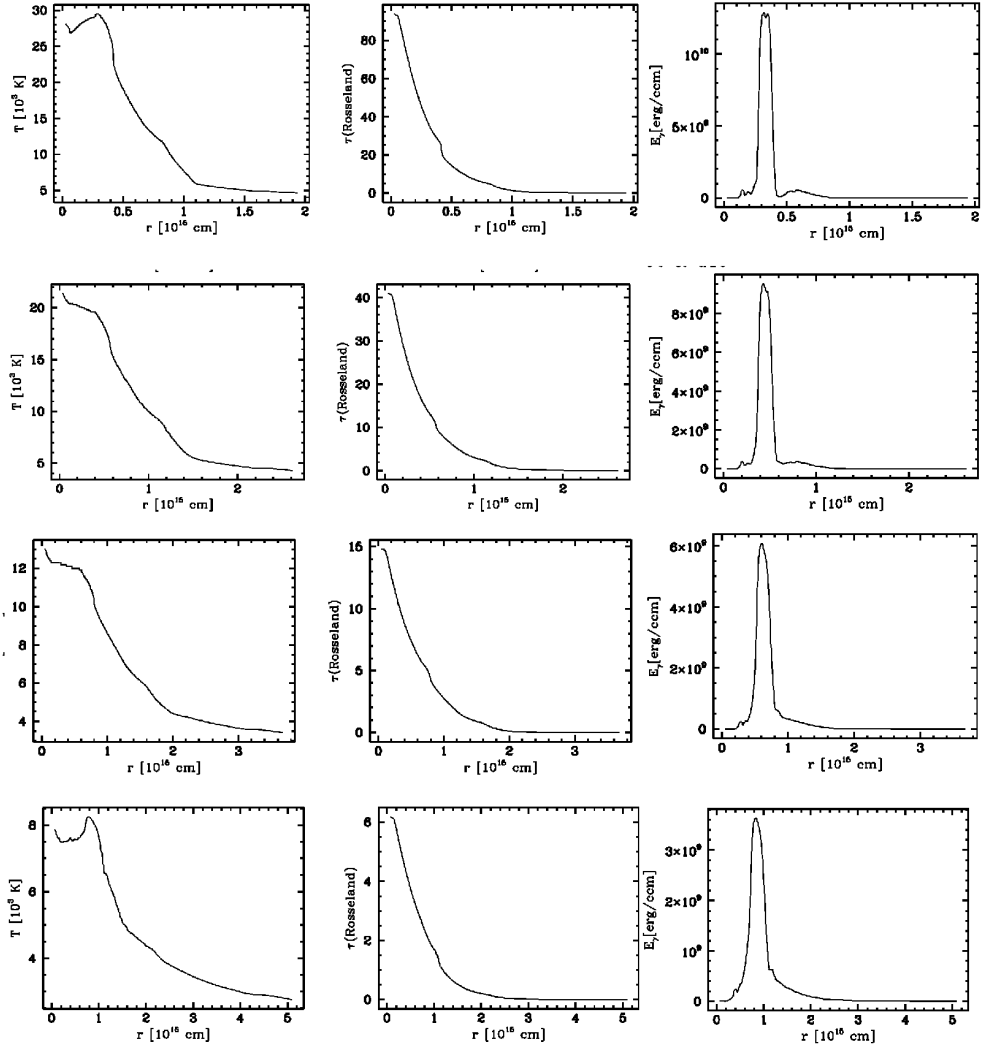}
%#\includegraphics[width=5.2cm,angle=270]{5p02822-9c.d11.factor1.ps}
%#\includegraphics[width=5.2cm,angle=270]{5p02822-9c.d15.factor1.ps}
%#\includegraphics[width=5.2cm,angle=270]{5p02822-9c.d22.factor1.ps}
%#\includegraphics[width=5.2cm,angle=270]{5p02822-9c.d29.factor1.ps}
\caption{Temperature (left), Rosseland optical depth (center), and energy 
deposition by $\gamma$-rays (right), as functions of radius at day 10, 14, 21 
and 28 (from top to bottom).}
\label{structures}
\end{figure}

Based on the density, temperature and chemical structure predicted from the
explosion model, $\gamma-ray$ transport and light curve calculations, we have 
constructed detailed NLTE-spectra for 11~d, 15~d, 22~d and 29~d after the 
explosion (days $-4$, 0, +7 and +14 after maximum light). For the subluminous 
model with the transition density 
$\rho_{tr}=8\times 10^6$ g~cm$^{-3}$ (Fig.~\ref{density}), the evolution of 
the temperature, energy deposition by 
$\gamma$-rays, and Rosseland optical depth is given as a function of radius 
(Fig.~\ref{structures}). Obviously, the total optical depth drops rapidly with
time from 98, 40, 15 to 6 due to the cooling; much faster than quadratic as 
would be expected from the geometric dilution alone (98, 52, 25 and 12), and 
thus more rapidly than in normal-bright SNe~Ia \citep{hoflich93,khokhlov93}). 
As mentioned above, the reason is the rapid drop in the 
mean opacity for $T \leq 6000 ... 8000 K$ which causes the outer layers to 
become almost transparent. Consequently, the photosphere recedes quickly in mass
and even starts to shrink in radius just a few days after the explosion. The 
corresponding expansion velocities at the photosphere are approximately 14000, 
10500, 6500 and 4000 km~s$^{-1}$ at the four epochs computed. 
 
 The low photospheric temperatures are due to the small amount of
$^{56}Ni$ and its low expansion velocity which also causes the local trapping of
$\gamma$-rays (Fig.~\ref{structures}). In contrast to normal SNe~Ia, most of 
the decay energy is deposited well below the photosphere up to about 2 weeks 
after maximum light. This explains, in part,  the low ionization stages seen in 
the IR-spectra (see below).
 
The location (and thus the velocity) of the photosphere determines which layers
in the supernova envelope will form the spectral features.  Before maximum 
light, the spectra sample the layers of explosive carbon burning which 
are O, Mg, Ne and Si-rich (Fig.~\ref{abundance}). Up to 2 weeks after maximum 
light, spectra are formed within layers of incomplete silicon burning which are 
Si, S and Ca-rich. Only thereafter are the layers which are dominated by iron 
group elements finally exposed. This is very different from models for normal 
luminosity SNe~Ia for which the iron rich layers are already exposed near 
maximum light.  These profound differences are key for our understanding of the 
optical and IR spectra of subluminous SNe~Ia and the differences compared to 
normal SNe~Ia. 
 
\begin{figure}
\includegraphics[width=10.2cm,angle=270]{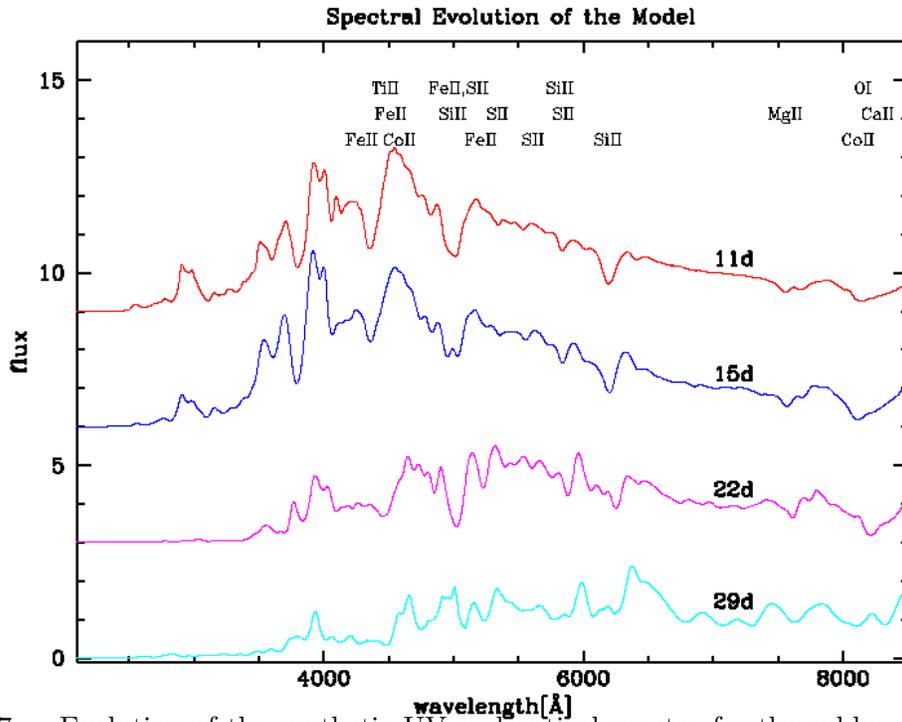}
\caption{
Evolution of the synthetic UV and optical spectra for the subluminous 
model 5p0z22.8 at 10.6, 15, 22  and 
29 days after the explosion. The flux has been normalized at  7000 \AA. }
\label{mod.opt}
\end{figure}

\begin{figure}
\includegraphics[width=10.2cm,angle=270]{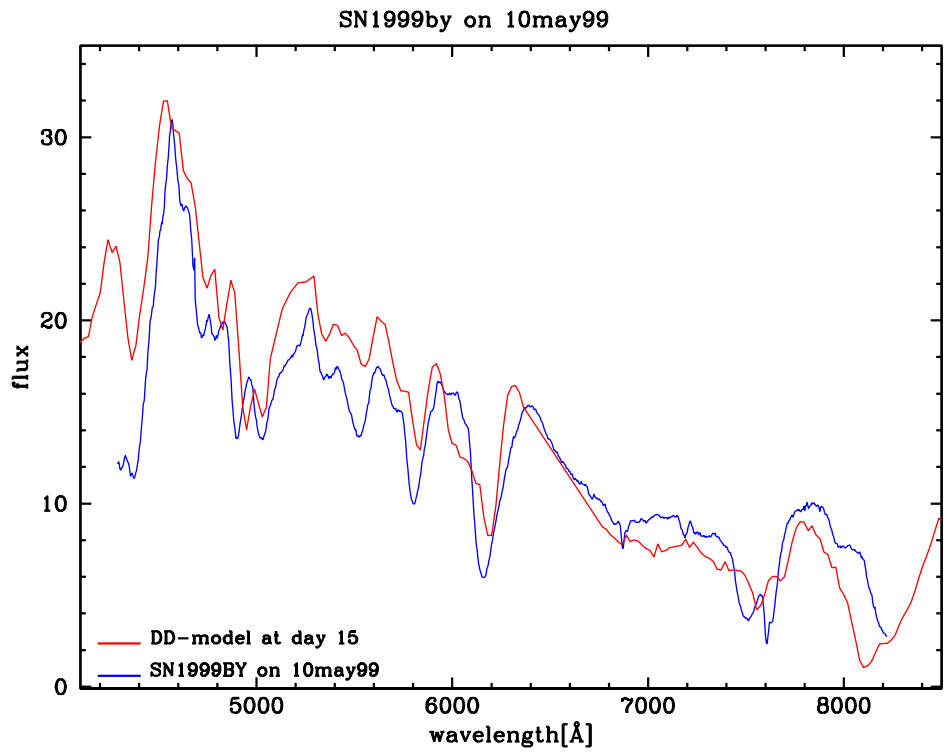}
\caption{Comparison of the optical spectrum of SN1999by on May 10, 1999, 
(blue) with the theoretical spectrum of 5p0z22.8 at maximum light (red,  
15 days after the explosion).}
\label{mod.opt2}
\end{figure}

Optical spectra are not the main subject of this paper, and detailed spectra
have not been available for this analysis except for maximum light 
(Fig.~\ref{mod.opt2}, \citealt{howell01}).     
However, for completeness, the time sequence of optical spectra is given in 
Figure~\ref{mod.opt}. Overall, the spectra show an evolution typical for SN~Ia, 
(e.g. SN~1994D, H95). They can be understood as a consequence of the declining 
temperature and total optical depth with time, and the rapidly dropping 
temperature. The Doppler shift of the absorption component decreases rapidly 
because of the receding photosphere. For subluminous models, the entire envelope
becomes almost transparent and the spectra start to enter the nebular phase at 
about two weeks after explosion. This marks the end of the applicability of our 
atomic models. Some features (see Fig.~\ref{mod.opt2})
are typical of subluminous SNe~Ia and are a 
consequence of the low luminosity and photospheric temperature around maximum 
light. The \ion{Si}{2} line at 5800 \AA\ is strong relative to the line at 6150 
\AA, and the flux near 4000 \AA\ was depressed as in other subluminous SNe, and
features due to \ion{Ti}{2} were strong as was the \ion{O}{1} line at 7500 \AA\
\citep{garnavich99}.  As already suggested by 
\citet{nugent97}, the strong \ion{Ti}{2} absorption and the relative strength of
the Si lines can be understood as a temperature effect. Within DD models with 
$ M_{Ch}$ progenitors, the strong \ion{O}{1} line is a direct consequence of 
the fact that the spectra at maximum light are formed in massive layers of 
explosive carbon burning. 

While the structures and synthetic spectra of the models 5p0z22.8 and 5p0z22.10 
are very similar, 5p0z22.8 is slightly preferred from a consideration of the 
expansion velocities at the photosphere which are 10500 and 11500 km s$^{-1}$, 
respectively.  Therefore, we will will adopt 5p0z22.8 in the following. Based on
parameterized LTE model atmospheres (SYNOW, \citealt{fisher99}), a detailed 
analysis of the Doppler shifts of lines has been provided by \citet{garnavich}. 
They found \ion{Si}{2} in the entire range between 11,300 down to 6500 
km~s$^{-1}$ which is consistent with our hydrodynamical models (see 
Fig.~\ref{structures}).
 
\subsection{Analysis of the IR-spectra of SN~1999by}
 
\begin{figure}
\includegraphics[width=10.2cm,angle=270]{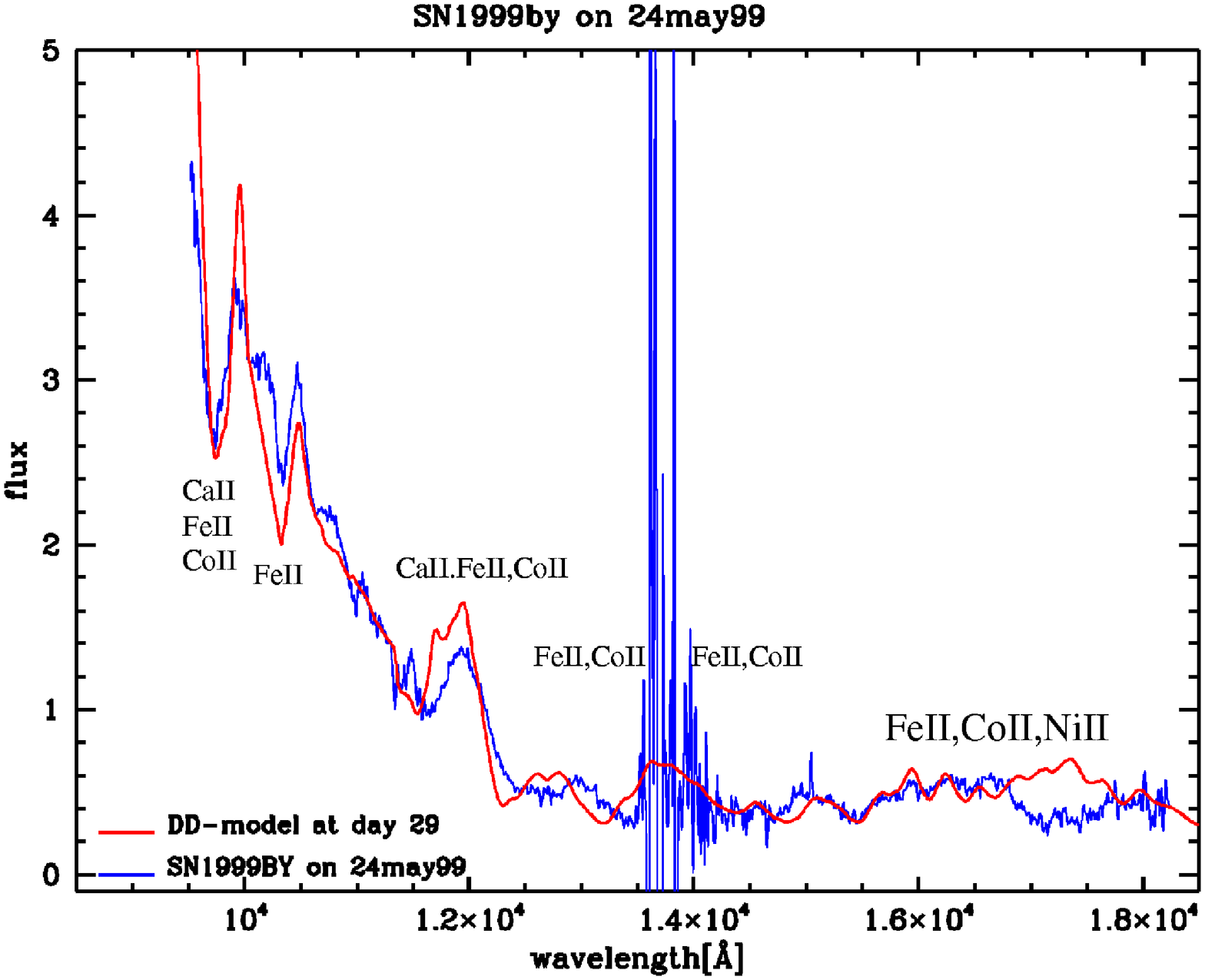}
\caption{Comparison of the observed NIR spectrum of SN1999by on May 24, 1999,
with the theoretical IR-spectrum of 5p0z22.8 14 days after maximum light
(29 days after the explosion). Between 1.35 and 1.45~\micron\ the S/N in the
observed spectrum is very low due to strong telluric absorption.}
\label{ir.may24}
\end{figure}

Near IR-spectra (0.9 to 2.5  $\mu$m) of 5p0z22.8 at day 11, 15, 22 and 29 after
explosion are compared to those of SN1999by at May 6, 10, 16, and 24 
(Figs.~ref{ir.may6}--\ref{ir.may24}). 
The spectra are characterized by overlapping P-Cygni and absorption lines, and
emission features. Unless noted otherwise, the observed wavelengths quoted
the emission component.
The overall energy distribution in the continuum is determined by 
bound-free and Thomson opacities and the temperature evolution.
In general, the spectra and their time evolution agree
reasonably well with the model predictions and all the major features can be 
identified.                                  
 
Few detailed analyses and line identifications of IR spectra have been
performed \citep{meikle96,bowers97,hoflich97,wheeler98,hernandez00} and these
were limited to spectra of SN~1994D, SN~1998bu and post-maximum spectra of 
SN1986G. To put our analysis given below in context with normal-luminosity 
SNe~Ia, we will refer to those works as reference points. 
 
\subsubsection{Early time IR-spectra}
  
\begin{figure}
\includegraphics[width=10.2cm,angle=270]{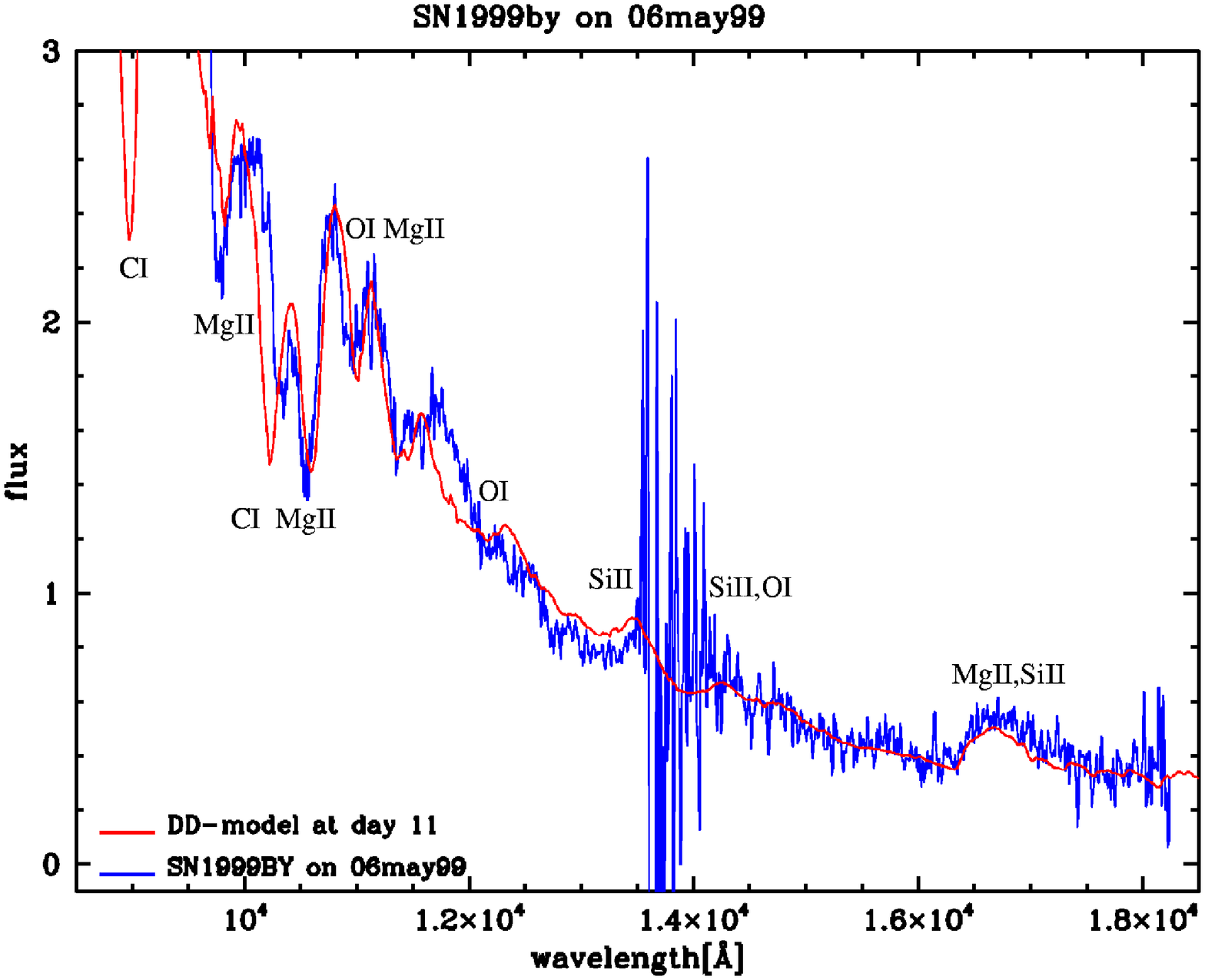}
\includegraphics[width=10.2cm,angle=270]{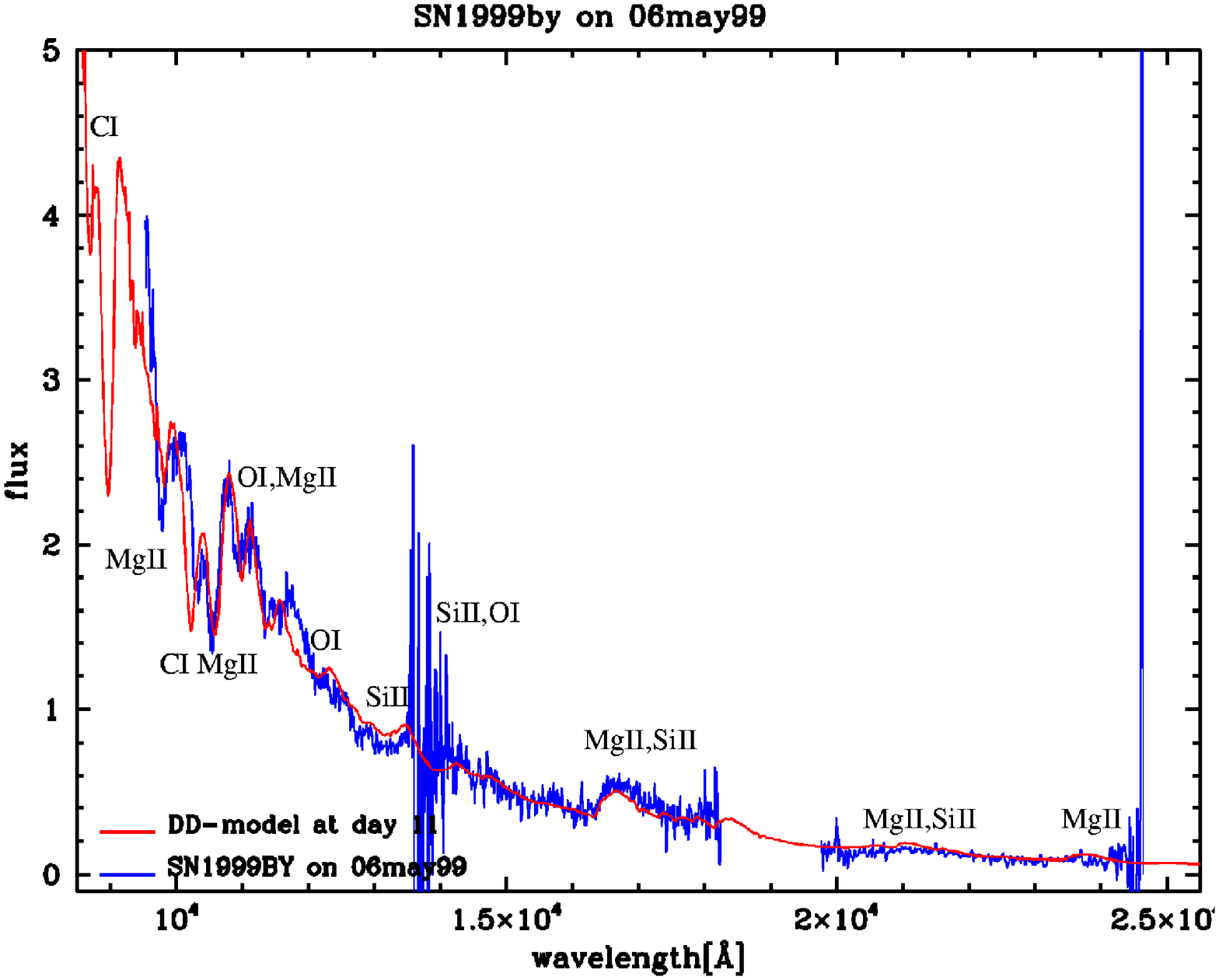}
\caption{Comparison of the observed NIR spectrum of SN1999by on May 6, 1999,
with the theoretical IR-spectrum of 5p0z22.8 $\approx 4$ days before maximum 
light (10 days after the explosion).  Between 1.35 and 1.45~\micron, and 
longward of 2.45~\micron, the S/N in the observed spectrum is very low due
to strong telluric absorption.} 
\label{ir.may6}
\end{figure}

\begin{figure}
\includegraphics[width=10.2cm,angle=270]{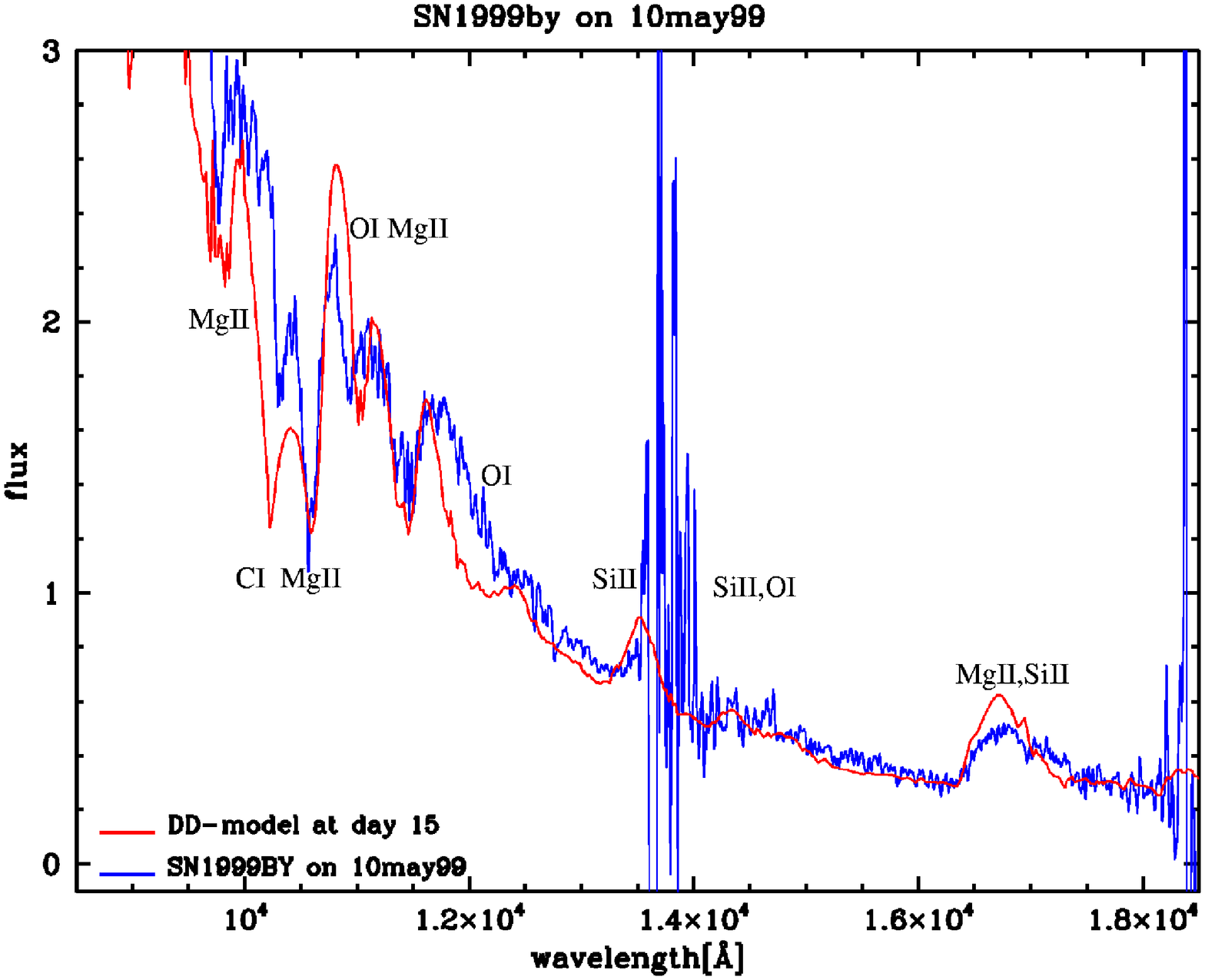}
\includegraphics[width=10.2cm,angle=270]{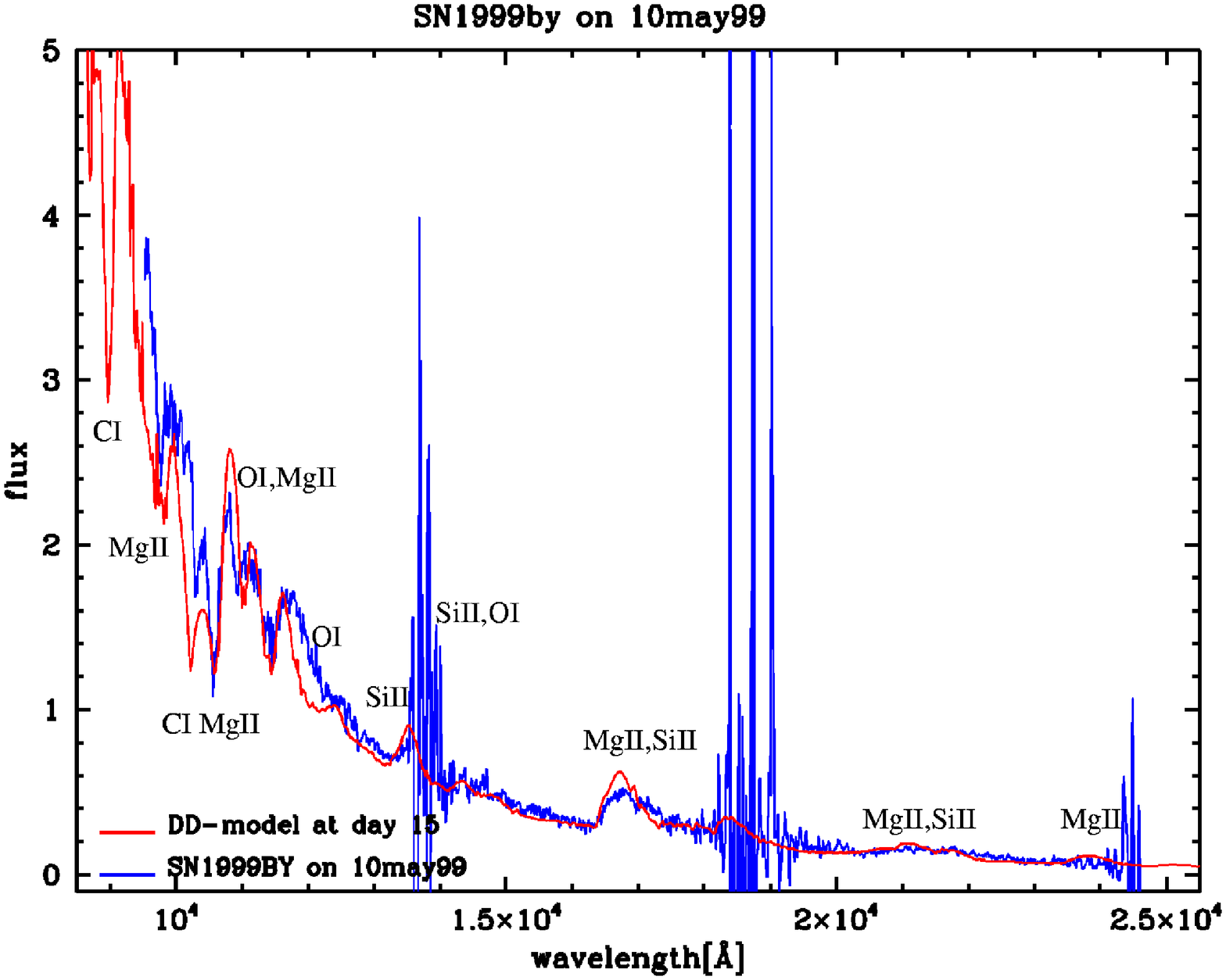}
\caption{Comparison of the observed NIR spectrum of SN1999by on May 10, 1999,
with the theoretical IR-spectrum of 5p0z22.8 at maximum light (15 days after
the explosion).  Between 1.35 and 1.45~\micron, between 1.8 and 1.9~\micron,
and longward of 2.45~\micron, the S/N in the observed spectrum is very low due
to strong telluric absorption.}
\label{ir.may10}
\end{figure}

The observed NIR spectra from May 6 and 10 (Figs.\ \ref{ir.may6} \& 
\ref{ir.may10}) are quite similar, and comparison with the model spectra
show that both of these spectra are formed in layers of explosive carbon 
burning. Spectral features can be attributed to \ion{C}{1}, \ion{O}{1}, 
\ion{Mg}{2} and \ion{Si}{2}. 
Typically, several thousand overlapping lines contribute to the overall 
opacity but few strong features leave their mark on the spectrum.
 
Some weak features at about 2.1 and 2.38 \micron\ are due to
\ion{Mg}{2} (21369, 21432, 24041, 24044, 24125 \AA) and \ion{Si}{2} (21920, 
21990 \AA).  The strong, broad feature between 1.62 and 1.75 \micron\ is a 
blend produced  by \ion{Mg}{2} lines (16760, 16799, 174119, 17717 \AA) and 
\ion{Si}{2} (16907, 16977, 17183, 17184 \AA). Its blue edge is determined by 
\ion{Mg}{2}. A similar feature is also present in normal luminosity SNe~Ia but 
\ion{Mg}{2} does not contribute because the corresponding spectra are formed in 
layers of explosive oxygen burning.                         
 
The strong feature that peaks near 1.35~\micron\ is due to \ion{Si}{2} 
(13650~\AA), and weaker features with peaks near 1.23~\micron\ and 1.43~\micron\
are a \ion{Si}{2} multiplet (13692 to 13696 \AA) and a blend of \ion{O}{1} 
(14542\AA) and \ion{Si}{2} (14454 \AA), respectively.  A very weak feature in 
the synthetic spectrum near 1.23\micron\ can be attributed to \ion{O}{1}
(13164 \AA), but is well below the noise level of the observed spectra.

Features with P-Cygni absorption minima at 1.12, 1.06, 1.03 and 1.0 \micron\ are
produced by \ion{Mg}{2} (11620, 11600 \AA), \ion{O}{1} (11302, 11286 \AA), 
\ion{Mg}{2} (10914, 10915   \& 10950   \AA), \ion{C}{1} (10683 \& 10691 \AA),
and \ion{Mg}{2} (10092 \AA), respectively. The multiplet of \ion{Mg}{2} 
1.09~\micron\ is also prominent in normal luminosity SNe~Ia but the other 
features are weak and blended with \ion{Fe}{2}, \ion{Co}{2} and \ion{Ni}{2}.
 
The observed spectra go down to about 9700 \AA. However, there are some 
interesting features at shorter wavelengths which are worth mentioning.  
Foremost, the model predicts a very strong feature at about 9100 \AA\ due to 
\ion{C}{1} (9405 \AA). Its velocity provides an important constraint on the 
minimum velocity for the unburned region, and possible mixing. Other features 
are produced by \ion{Mg}{2} (9344 \AA), \ion{Ca}{2} (IR-triplet), and \ion{O}{1}
(8446 \AA) and \ion{Mg}{2} (8246 \AA).
 
Despite the overall agreement, some discrepancies remain. First, the \ion{C}{1}
line at 10691\AA\ is somewhat too strong, and its Doppler shift on May 6 is too 
large by about 1500 km~s$^{-1}$. This feature is formed well above the 
photosphere (13000 km~s$^{-1}$). This might indicate that SN~1999by has a 
slightly lower temperature in the outer layers compared to the model, which 
would cause less excitation of carbon and, might cut off the high velocity 
contribution to the absorption. Secondly, in the model spectrum there seems to 
be a lack of emission produced by \ion{Si}{2} near 1.18 \micron. We note that
similar problems with Si are also commonly present in the optical \ion{Si}{2} 
line at about 6380 \AA \ for normal SNe~Ia \citep{hoflich95,lentz00}. They may 
be related to the excitation process of Si but their origin is not well 
understood.
 
\subsubsection{IR-spectra after maximum light}

\begin{figure}
\includegraphics[width=10.2cm,angle=270]{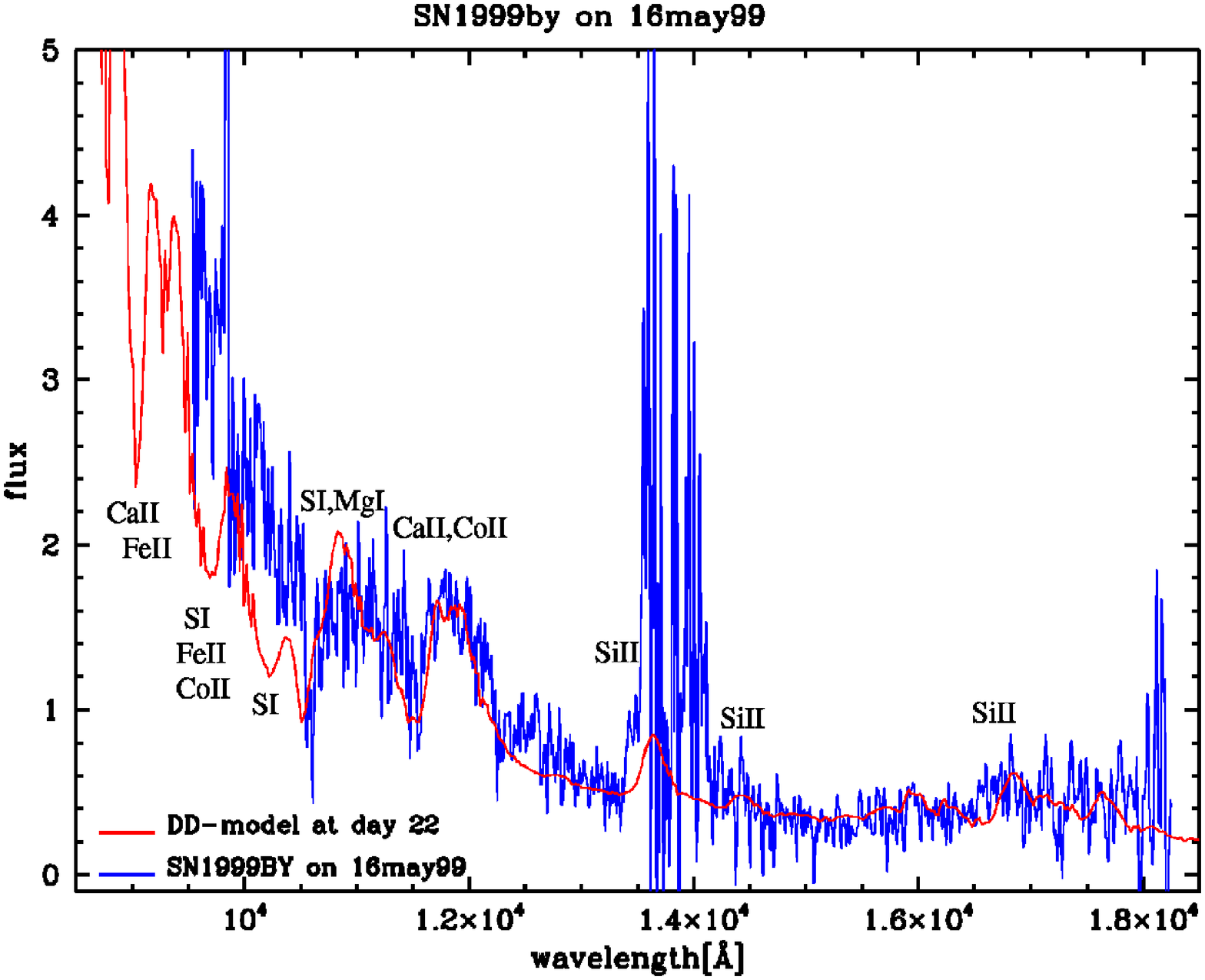}
\caption{Comparison of the observed NIR spectrum of SN1999by on May 16, 1999,
with the theoretical IR-spectrum of 5p0z22.8 seven days after maximum light 
(22 days after the explosion).  Between 1.35 and 1.45~\micron\ the S/N in the 
observed spectrum is very low due to strong telluric absorption.}
\label{ir.may16} 
\end{figure}

The spectrum from May 16 (Fig.~\ref{ir.may16}) is formed in layers of 
explosive oxygen burning. The lines can be attributed to \ion{Mg}{1},
\ion{Ca}{2}, \ion{Si}{2} and \ion{S}{1}.
All features due to \ion{Ca}{2} and \ion{Si}{2} seen in the 
pre-maximum spectra can still be identified. No lines due to \ion{O}{1} and 
\ion{C}{1} are present, but strong lines of \ion{Ca}{2} and \ion{S}{1} appear. 
Strong features at about 1.14 \micron\ and 1.1 \micron\ are due to 
\ion{Ca}{2} lines (11784, 11839 \& 11849 \AA) and \ion{S}{1} (10821  \AA) \& \ion{Mg}{1} (11800 \AA), and the broad feature at 
0.91~\micron\ (not covered in this spectrum, but included for completeness) 
is a blend of \ion{Ca}{2} (9890 \AA) and \ion{Mg}{2} (10009\AA). Weaker lines 
(\ion{Fe}{2}, 9997\AA;  \ion{S}{1}, 10455 \AA) contribute to blends 
around 1.03 \micron.
 
Some weak lines due to \ion{Co}{2} and \ion{Fe}{2} lines occur throughout the 
spectrum but the most characteristic feature in post maximum IR-spectra of 
normal SNe~Ia is missing. Starting a few days after maximum, normal SNe~Ia show 
a wide emission feature between 1.5 and 1.8 \micron\ produced by thousands of 
\ion{Fe}{2}, \ion{Co}{2} \& \ion{Ni}{2} lines.  Typically, their emission flux 
is about twice as large as the adjoining continuum. This feature marks the lower
velocity end of the region of complete silicon burning, and it seems to be 
common in all normal luminosity SNe~Ia such as SN~1986G, SN~2000br and SN~2000cx
(\citealt{wheeler98}; Marion 2001, priv. comm.; \citealt{rudy01}). In 5p0z22.8, this broad
emission feature is not seen because the photosphere has yet to recede to the 
layers of complete silicon burning.  On May 24th (Fig.~\ref{ir.may24}) this 
feature finally appears but is very weak.    Overall, two weeks after maximum 
light, the spectrum in dominated by a large number of \ion{Fe}{2} and 
\ion{Co}{2} lines. Strong features of \ion{Fe}{2} are around 9600 \AA\, 1.2 
$\mu$m and 1.4 $\mu$m with strong \ion{Co}{2} lines from 9500 to 12443 \AA, 
and 14000\AA\ -- 16700 \AA.

In our synthetic spectrum the \ion{Fe}{2}, \ion{Co}{2} \& \ion{Ni}{2} feature
1.6 to 1.8  $\mu$m does not show a local minimum at $\approx 1.7 \mu m$ as 
observed in SN1999by, and as also predicted by  normal bright models for SN1986G
\citep{hoflich97,wheeler98}.  In our subluminous model, the emission near 
1.7~\micron\ comes from the central Ni-rich region. The absence of such emission
in the observed spectrum may indicate that the very central region 
($M(r)\leq 0.2 M_\odot$) of the model is too transparent compared SN1999by. 
This discrepancy is not critical for measuring the chemical structure of the 
envelope because this is determined by the blue edge. The difference in opacity 
may be caused by missing iron lines in the line list (\S 3.1.3), or an 
underestimation of the excitation by non-thermal electrons because we assume 
local deposition for the electrons from the $\beta^+$ decay of $^{56}Co$.
 
\subsection{Mixing processes}

As we have seen, model 5p0z22.8 reproduces the basic features of both 
optical and IR spectra, including the time evolution.  This suggests that
DD models are a reasonably good description for subluminous SNe~Ia.
  
Unfortunately, we are still missing a  handle on an important piece of physics. 
The propagation of a detonation front is well understood but the description of 
the deflagration front and the deflagration to detonation transition (DDT) pose 
problems. Currently, state of the art allows us to follow the front only through
the phase of linear instabilities, i.e. only the early part of the deflagration 
phase. The resulting structures of these calculations cannot account for the 
observations of typical SNe~Ia (Khokhlov 2001), because a significant fraction 
of the WD ($\approx 0.5 M_\odot$) remains unburned.

Nevertheless, these calculations point toward a more general 
problem, or toward a better understanding of the nature of subluminous SNe~Ia.
On a microscopic scale, a deflagration propagates due to heat conduction by 
electrons. Though the laminar flame speed in SNe~Ia is well known, the front has
been found to be Rayleigh-Taylor (RT) unstable, increasing the effective speed 
of the burning front \citep{nomoto76}. More recently, significant
progress has been made toward a better understanding of the physics of flames. 
Starting from static WDs, hydrodynamic calculations of the deflagration fronts 
have been performed in 2-D \citep{niemeyer95,lisewski00}, and full 3-D 
\citep{khokhlov95,khokhlov01}.  

The calculations by \citet{khokhlov01} 
demonstrated a complicated morphology 
for the burning front. Plumes of burned material will fill a significant 
fraction of the WD, and unburned or partially-burned material can be seen near 
the center. Thus iron-rich elements will not be confined to the central region 
as in 1-D models.  \citet{khokhlov01} finds that while the expansion of the 
envelope becomes almost spherical, the inhomogeneous chemical structure will 
fill about 50 to 70\% of the star (in mass). If a DDT occurs at densities 
needed to reproduce normal-bright SNe~Ia, most of the unburned fuel in these 
regions will be burned to iron-group elements during the detonation phase. This 
will eliminate the chemical inhomogeneities. However, in our subluminous model, 
the chemical structure imposed during the deflagration phase must be expected 
to survive because no or very little $^{56}Ni$ is produced during the 
detonation. 
 
\begin{figure}
\includegraphics[width=10.2cm,angle=270]{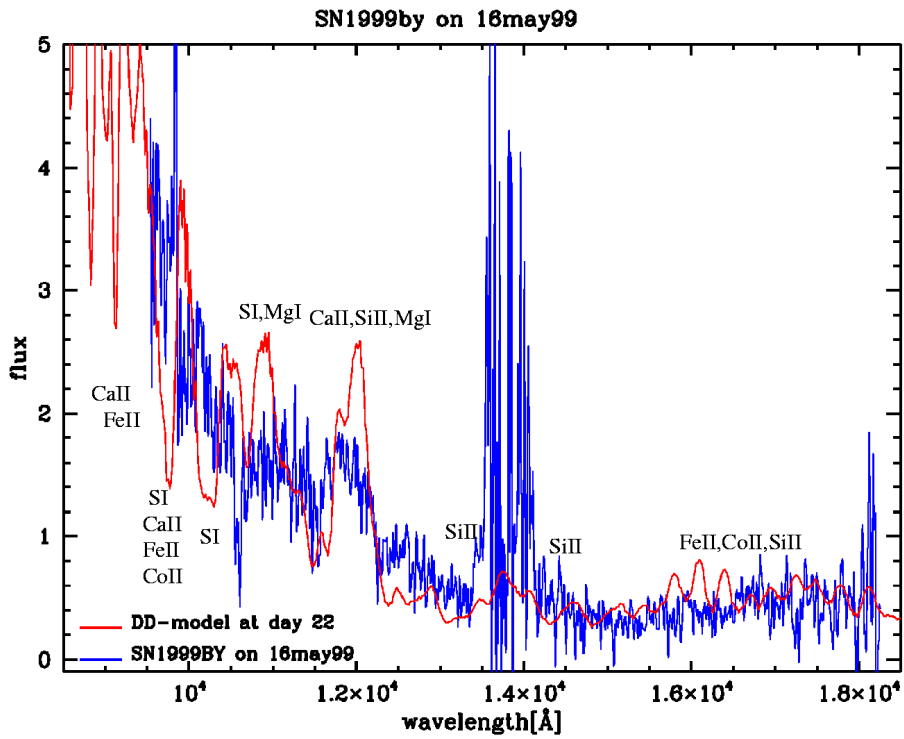}
\caption{Comparison of the observed NIR spectrum of SN1999by on May 16, 1999,
with the theoretical IR-spectrum of 5p0z22.8 seven days after maximum light if 
we impose mixing in the inner 0.7 $M_\odot$.  Line blanketing due to
a large number of FeII and CoII lines becomes strong as on May 24, 1999 in the 
unmixed model (see Fig.~\ref{ir.may16}).}
\label{mix.may16}
\end{figure}

In order to test for this effect, we have mixed the inner layers of model 
5p0z22.8 up to an expansion velocity of 8000 km~s$^{-1}$. This brings iron-group
elements into the outer regions and mixes intermediate mass elements (Si, S, Ca 
and some O) into the inner, hotter layers. The layers with expansion velocities 
less than 8000 km~s$^{-1}$ become visible a few days after maximum light (see 
above).  In Figure~\ref{mix.may16}, a comparison is shown between the spectrum 
of SN1999by on May 16th, 1999, and the synthetic spectrum about one week after 
maximum light, and problems with the spectral fit are apparent. As can be 
expected from the discussion above, the effective extension of the iron-rich 
layers to 8000 km~s$^{-1}$ causes the 1.5 to 1.8 \micron\ blend to become rather
strong, and line blanketing due to iron-group elements (\ion{Fe}{2} \& 
\ion{Co}{2}) become too strong in the region around 1 to 1.2 \micron. 
Moreover, at about 0.99, 1.02, 1.05 \& 1.16 \micron, (absorption components) 
strong blends can be seen due to \ion{O}{1} (11302, 11287 \& 11297\AA), 
\ion{Si}{2} (11715, 11745, 13650 \& 13650 \AA), \ion{Ca}{2} (9890,
9931, 9997, 11839 \& 11745 \AA) and \ion{S}{1} (10455 \& 108212 \AA).  

We have 
to take into account that our approach assumes microscopic mixing whereas RT    
instabilities provide large scale inhomogeneities. If $^{56}Ni$ and the 
intermediate mass elements are separated, direct excitation of intermediate mass
elements may be reduced. With respect to the radiation transport, the main 
difference is the covering factor. This will effect line blanketing and thus 
the line blanketing between 1 to 1.2 $\mu$m may not pose a problem. However, the
emission features between 1.5 to 1.8 $\mu$m will not go away. The reduced 
excitation by non-thermal electrons and $\gamma$'s in case of macroscopic mixing
are important at low optical depths. Less excitation may reduce the problem with
the O, Si, S \& Ca lines but the problem will no disappear because the 
corresponding features are formed close to the photosphere where the excitation 
is thermal in nature.
  
These problems strongly suggest that large scale mixing did not occur in 
SN~1999by.  This may provide a key for our understanding of the nature of these 
objects.

\section{Implications for the SN~Ia progenitors, and alternative scenarios}

As previously discussed, the pre-conditioning of the WD
before the explosion may provide a key for understanding the nature of 
subluminous SNe~Ia. Such pre-conditioning may include the main sequence mass of 
the progenitor mass and its metallicity \citep{dominguez01,hoflich98,iwamoto99},
the accretion history \citep{langer00}, or large-scale velocity fields such as 
turbulence prior to the runaway (\citealt{hs01}). 
Up to now, none of these effects have been included in detailed 3-D 
calculations. Therefore, the suggestions (below) require further investigation 
before taken as explanation.  

Beyond its subluminous nature, SN~1999by showed another peculiarity with respect
to normal-bright SNe~Ia. The observed polarization of SN~1999by was rather high
($P \approx 0.7 \%$) with a well-defined axis of symmetry, whereas normal SNe~Ia
show little or no polarization ($P \leq 0.2 \%$; \citealt{wang01}).
>From a detailed analysis of the polarization spectra, \citet{howell01} concluded
that the overall geometry of SN~1999by showed a large scale, probably rotational
asymmetry of about 20\%. The authors suggested several explanations. Either, 
SN~1999by may be the result of the explosion of a rapidly rotating WD close to 
the break-up, or a result of a merger scenario. In a rapidly rotating WD, 
large-scale circulations \citep{erigichi85} might influence the deflagration 
phase of explosive burning, perhaps by breaking up large eddies and 
preventing their rise.  

Alternatively, it should be remembered that the chemical
structure of the exploding WD depends on the pre-expansion of the WD prior to 
the detonation phase. More precisely, we need to burn about $0.2 M_\odot$ to 
lift the WD in its gravitational potential. This mass is significantly more than
is burned during the smoldering phase (the phase of slow convective 
burning just prior to the explosion). Even in the presence of strong turbulence,
the amount of burned material is much less ($\leq 10^{-2} M_\odot$, 
\citealt{hs01}).  In all models for the ignition (e.g. 
\citealt{nomoto82,garcia95}), the central temperature rises by non-explosive 
carbon burning during the last few minutes to hours 
before the runaway because the heat cannot be dispersed. It may be feasible that
large scale velocity fields due to rotation may extend this phase, and 
significantly increase the amount of burning prior to the explosion. Thus, the 
pre-expansion in subluminous SNe~Ia may occur directly during the smoldering 
phase, and the deflagration phase may be even skipped, i.e. a detonation occurs 
promptly, and we may have a smoldering detonation model (SD). In SD-models, no $^{56}Ni$
is produced during the phase of pre-expansion and, consequently, this mechanism does
not predict a lower limit for $M_{Ni}$ and, thus, $M_V$.  In principle, SD-models may be
an attractive alternative also for normal-bright SNe because they omit the need for a DDT but,
from the present understanding of the runaway, in general, extensive burning prior to the
runaway cannot be expected (see above).
 From the point of the spectral analysis, this option is intriguing 
because burning during the smoldering phase precedes only to O and the time 
scales of burning are sufficient to completely homogenize the chemical structure
of the WD.  

Finally, there is the question of alternative scenarios for SNe~Ia in the 
context of subluminous SNe~Ia. Pure deflagrations of $M_{Ch}$-WDs without any 
DDT may produce little $ ^{56}Ni$ but, according to current simulations, a large
fraction of the WD would remain unburned. Even if this problem were solved 
in the future, the problem of inhomogeneous chemistry and Si-rich layers close 
to the center would remain.  Helium triggered detonations of sub-Chandrasekhar 
mass WDs have been suggested in the past. However, their distinct feature is an 
outer layer of $^{56}Ni$ above a low mass layer of explosive oxygen burning, a 
layer of incomplete Si burning, and some $^{56}Ni$ close to the center. This 
chemical structure is in strong contradiction to the optical spectra and LCs 
of SN~1999by, and their evolution with time.
 
Alternatively, setting aside the problem of triggering the explosion, central 
detonations in low-mass WDs have been suggested in the past 
(e.g., \citealt{ruizlapuente93}). To produce little $^{56}Ni$, the explosion 
must occur in WDs with central densities $\leq 1 \times 10^7$ g cm$^{-3}$ which 
corresponds to $M(WD) \leq 0.7 M_\odot$. In contrast to the DD-models, these 
models would not show a lower limit for the $^{56}Ni$ production and $M_V$, 
and they would avoid the problem of mixing during a deflagration phase. 
Nonetheless, we regard these models as unlikely candidates for SN~1999by because
there are a couple of severe problems beside the lack of a triggering mechanism 
for the explosion.  The total mass of the ejecta is lower (by a factor of 2) 
than is required to keep the line forming region in the layers of explosive 
carbon and incomplete Si-burning up to two and four weeks after the explosion, 
respectively. The specific production of nuclear energy does not depend on 
$M_{WD}$. In this case, the diffusion time scales and, thus, the rise times to 
maximum light are $ \propto M_{WD}$ \citep[e.g.][]{pinto00}. Compared to the LCs
for a WD with $M_{Ch}$, we expect rise times to be shorter by $\approx 2$ which 
is incompatible with SN~1999by observations.
   
This leaves the merging scenario as an alternative as we may expect a 
significant asphericity in the explosion if the merging occurs on hydrodynamical
time scales.  Nevertheless, this alternative is not a very attractive one. Based
on 1-D parameterized models, \citet{khokhlov93} provided chemical structures for
explosions which may resemble merger models. As a general feature, these models 
show a shell-like structure with a massive outer layer of unburned C/O which 
encloses a thin, low mass layer of explosive carbon burning.  This pattern is
not consistent with the IR-spectra for SNe~1999by up to maximum light. However, 
it should be kept in mind that, besides the limitations of 1-D 
models for mergers, no chemical structures are available for merging subluminous
models.  A more severe problem for mergers may be that merger models tend to 
ignite prematurely and produce a detonation wave that burns the C/O WD to a 
O/Ne/Mg WD prior to the completion of the merger. Thus, merging may result in an
accretion induced collapse rather than a SNe~Ia \citep{saio98}.

\section{Summary}

IR-spectra of the subluminous SN1999by have been presented which cover the time
evolution from about 4 days before to 2 weeks before maximum light. This is the
first subluminous SN Ia (and arguably the first SN~Ia) for which IR spectra with
good time coverage are available. These observations allowed us to determine
the chemical structure of the SN envelope.

Based on a delayed detonation model, a self-consistent set of hydrodynamic 
explosions, light curves, and synthetic spectra have been calculated. This 
analysis has only two free parameters: the initial structure of the progenitor 
and the description of the nuclear burning front. The light curves and spectral 
evolution follow directly from the explosion model without any further tuning, 
thus providing a tight link between the model physics and the predicted
observables.
 
By varying a single parameter, the transition density at which detonation
occurs, a set of models has been constructed which spans the observed
brightness variation of Type~Ia supernovae.  The absolute maximum brightness 
depends primarily on the $^{56}Ni$ production which, for DD-models, depends 
mainly on the transition density $\rho_{tr}$.  The brightness-decline relation 
$M_V (\Delta M_{\Delta t =15d})$ observed in SNe~Ia is also reproduced in these
models.  In the DD scenario this relation is a result of the temperature 
dependence of the opacity, or more precisely, as a consequence of the rapid drop
in the opacities for temperatures less than about 7000 to 8000 K.

Within $M_{Ch}$ WD models, a certain amount of burning during the deflagration 
phase is needed to pre-expand the WD and avoid burning the entire star to 
$^{56}Ni$. This implies a lower limit for the $^{56}Ni$ mass of about $0.1 
M_\odot$ $\pm 25 \%$ and, consequently, implies a minimum brightness for SNe~Ia 
within this scenario.  The best model for SN~1999by 
($\rho_{tr} = 8 \times 10^7$~g~cm$^{-3}$, selected by matching to the predicted 
and observed optical light curves) is close to this minimum Ni yield.
The data are consistent with little or no interstellar reddening 
($E(B-V) \leq 0.12^m$), and the derived distance is $11 \pm 2.5$~Mpc 
or $12 \pm 1$~Mpc if we take the limit for $E(B-V)$ from the models or assume
$E(B-V)=0.015^m$ according \citet{schlegel98} for the galaxy.

Without any further modification, this subluminous model has been used to 
analyse the IR-spectra from May 6, May 10, May 16 and May 24, 1999, which 
correspond to $-4$~d, 0~d, +7~d and +14~d after maximum light. The photosphere
($\tau_{Thomson} = 1$) recedes from 14000, to 10500, 6500 and 4000 km~s$^{-1}$.
The observed and theoretical spectra agree reasonably well with respect to the
Doppler shift of lines and all strong features could be identified.

Before maximum light, the spectra are dominated by products of about
explosive carbon burning (O, Mg), and Si. Spectra taken at +7~d and +14~d
after maximum are dominated by products of incomplete Si burning.
At about 2 weeks after maximum, the iron-group elements begin to show up.
The long duration of the phases dominated by layers of explosive carbon burning 
and incomplete Si burning implies massive layers of these burning stages that 
are comparable with our model results.  ($\approx 0.45 $ and $ 0.65 M_\odot $ 
for explosive carbon and incomplete silicon burning, respectively.)  
This, together with the $^{56}Ni$ mass, argues that SN~1999by was the explosion 
of a WD at or near the Chandrasekhar mass.

Finally, we note that the observed IR spectra are at odds with recent 3-D 
calculations \citep{khokhlov01} which predict that large scale chemical 
inhomogeneities filling 50 -- 70\% (in mass) of the WD will be formed during the
deflagration phase.  When the effect of such inhomogeneous mixing is tested by 
mixing the inner layers of our SN~1999by model, significant differences appear 
between the model spectra and the observed data.  This suggests that no
significant large-scale mixing took place in SN~1999by.  The lack of observed 
mixing and the asphericity seen in SN~1999by may be important clues into the 
nature of subluminous SNe~Ia, and may be related to the reason for a low DDT 
transition density (and hence the low luminosity). Alternatively, we may have 
an extended smoldering phase of the WD prior to the explosion which skips the 
deflagration phase altogether (see \S 5). In either case, the pre-conditioning 
of the WD prior to the explosion seems to be a key for understanding SNe~Ia.

\acknowledgments
We would like to thank Paul Martini and Adam Steed for aiding us with data 
acquisition and J. C. Wheeler for carefully reading the manuscript and providing
helpful suggestions.  P. H. would like to thank NASA for its support by NASA 
grant NAG5-7937.  C. L. G. and R. A. F.'s research is supported by NSF grant 
98-76703.  The calculations for the explosion, light curves and spectra were 
performed on a Beowulf-cluster financed by the John W. Cox -- Fund of the 
Department of Astronomy at the University of Texas.

\begin{deluxetable}{ccccc}
\tablewidth{0pt}
\tablecaption{Log of Near-Infrared Observations\label{nirlog}}
\tablehead{
\colhead{Date} & \colhead{Epoch} & \colhead{Setup} & \colhead{Exp. Time} &
\colhead{Observer}}
\startdata
6 May	& -4 d	& 0.96--1.8~\micron\	& 4$\times$600~s	& Gerardy \\*
''	& ''	& 1.95--2.4~\micron\	& 3$\times$600~s	& '' \\
7 May	& -3 d	& 0.96--1.8 \micron\	& 3$\times$600~s	& ''\\*
''	& ''	& 1.95--2.4 \micron\	& 3$\times$600~s	& ''\\
8 May	& -2 d	& 0.96--1.8 \micron\	& 3$\times$600~s	& ''\\
9 May	& -1 d	& 0.96--1.8 \micron\	& 3$\times$600~s	& ''\\*
''	& ''	& 1.2--2.2 \micron\	& 3$\times$600~s	& ''\\
10 May	& 0 d	& 0.96--1.8 \micron\    & 3$\times$600~s        & ''\\*
''      & ''    & 1.2--2.2 \micron\     & 3$\times$600~s        & ''\\*
''      & ''    & 1.95--2.4 \micron\    & 3$\times$600~s        & ''\\
16 May	& 6 d	& 0.96--1.8 \micron\    & 3$\times$600~s   & Martini \& Steed\\
18 May	& 8 d	& 0.96--1.8 \micron\    & 3$\times$600~s   & ''\\
20 May	& 10 d	& 0.96--1.8 \micron\    & 3$\times$600~s   & ''\\
21 May	& 11 d	& 0.96--1.8 \micron\    & 3$\times$600~s   & Sakai \\
24 May	& 14 d	& 0.96--1.8 \micron\    & 3$\times$600~s   & '' \\
\cutinhead{NIR Photometry}
Date	& Epoch	& J	& H	& K\\
\cline {1-5}\\
6 May	& -4 d	& $13.28 \pm 0.09$ & $13.23 \pm 0.05$ & $13.19 \pm 0.05$ \\
7 May	& -3 d  & $13.28 \pm 0.09$ & $13.21 \pm 0.05$ & $13.14 \pm 0.05$ \\
8 May	& -2 d	& $13.07 \pm 0.09$ & $13.04 \pm 0.05$ & \nodata 
\enddata
\end{deluxetable}

\begin{deluxetable}{lccccccc}
\tablewidth{0pt}
\tablecaption{Properties of the models\label{modelprop}}
\tablehead {
\colhead {Name} &
\colhead {$\rho_{tr} [10^6 g/cm^{-3}]$}&
\colhead {$^{56}Ni~ [M_\odot]$}&
\colhead {$M_V$} &
\colhead {$t_V$} &
\colhead {$B-V  $}&
\colhead {$\Delta M_{15}$ }}
\startdata
5p0z22.8 &  8.  &    0.095 &   -17.21 &  13.9   &   0.47 & 1.45 \\
5p0z22.10 &  10. &    0.107 &   -17.35 &  14.1   &   0.42 & 1.37 \\
5p0z22.12 &  12. &    0.137 &   -17.63 &  14.7   &   0.38 & 1.32 \\
5p0z22.14 &  14. &    0.153 &   -17.72 &  14.9   &   0.22 & 1.30 \\
5p0z22.16 &  16. &    0.268 &   -18.72 &  15.8   &   0.14 & 1.26 \\
5p0z22.18 &  18. &    0.365 &   -18.82 &  16.6   &   0.08 & 1.21 \\
5p0z22.20 &  20. &    0.454 &   -18.96 &  17.0   &   0.02 & 1.19 \\
5p0z22.23 &  23. &    0.561 &   -19.21 &  18.2   &  -0.02 & 1.05 \\
5p0z22.25 &  25. &    0.602 &   -19.29 &  18.6   &  -0.02 & 1.00 \\
5p0z22.27 &  27. &    0.629 &   -19.35 &  19.0   &  -0.03 & 0.91 \\
\enddata
\end{deluxetable}

\clearpage

\clearpage
 
\clearpage
 
\clearpage

\clearpage
\clearpage

\clearpage

\clearpage
 
\clearpage
 
\clearpage

\clearpage

\clearpage

\clearpage

\clearpage

\clearpage 

\end{document}